\documentclass[%
 aps,%
 amsmath,amssymb,
 showkeys,
]{revtex4-2}
\usepackage{graphicx} 
\usepackage{subcaption}
\usepackage{hyperref}
\usepackage{multirow}
\usepackage{booktabs}
\usepackage{soul}
\usepackage[table]{xcolor}
\usepackage[dvipsnames]{xcolor}
\usepackage{natbib}
\setcitestyle{authoryear,open={(},close={)},sort}

\newcommand{\Ham}{\mathcal{H}}
\newcommand{\DefFigWidth}{0.47\textwidth}

\begin{abstract}
Using tracer particles propagating in the environment provided by general relativistic magnetohydrodynamic
simulations, this paper addresses fundamental questions related to the confinement and acceleration properties 
in accretion disks around compact objects. 
By varying the spin parameter of the central black hole and the accretion regime of the disk, a comprehensive 
picture is obtained for particles whose Larmor radius is comparable to the gravitational radius of the system.
It is found that such particles can escape the system at the outermost boundary of the simulations, though, 
depending on the accretion regime, the time it takes them to escape, and the energetic dynamics along the 
trajectory can vary significantly. Dynamics near the black hole are even more sensitive to environmental 
details as black holes with higher spin values generate stronger magnetic fields, which influence particle 
motion in non-trivial ways. 
The primary acceleration mechanism is identified as turbulent scattering, with the jet-sheath emerging as 
the dominant site for particle energization
\end{abstract}

\begin{document}

\title{Energetic Particle Confinement In Black Hole Accretion Disks}

\author{D. Villa}
 \affiliation{Max Planck Institute for Plasma Physics, Boltzmannstraße 2, 85748, Garching, Germany}
 \email{daniele.villa@ipp.mpg.de}

\author{F. M. Rieger}
 \affiliation{Max Planck Institute for Plasma Physics, Boltzmannstraße 2, 85748, Garching, Germany}
 
\author{F. Jenko}
 \affiliation{Max Planck Institute for Plasma Physics, Boltzmannstraße 2, 85748, Garching, Germany}

\author{J. M. Stone}
  \affiliation{Department of Astrophysical Sciences, Princeton University, Princeton, NJ 08544, USA}
\affiliation{School of Natural Sciences, Institute for Advanced Studies, 1 Einstein Drive, Princeton, NJ 08540, USA}
 
\date{\today}

\maketitle

\section{Introduction} 
The environment surrounding massive black holes (BHs) is a leading candidate site for the acceleration 
of energetic charged particles, as revealed by their multi-messenger signatures 
\citep[e.g.,][]{boulanger2018imagine,rieger2018,de2022physics,globus2025}. A recent detection of particular 
interest was reported by the IceCube Collaboration \citep{icecube2022evidence}, which identified an excess 
in the neutrino flux in the energy range $(1-15)$ TeV originating from a region of the sky coincident with
the Seyfert II active galactic nucleus (AGN) NGC~1068 \citep{bland1997ringberg, padovani2024high}.  
Analysis of this source \citep{padovani2024high}, in particular with regard to its bright X-ray emission but 
weak gamma-ray counterpart, suggests that a compact ($r \lesssim 30 \; r_g$)  region around the BH, commonly 
referred to as the ``corona'', is the most likely origin of the  observed neutrino flux.\\
The overall properties of such a region can be constrained through phenomenological arguments, but the 
details of its composition remain uncertain, with a growing body  of literature exploring these unresolved 
aspects \citep[e.g.,][]{murase2022,eichmann2022solving,fiorillo2024,das2024,padovani2024high,lemoine2025neutrinos}.  
Given the experimental challenges with directly probing such environments, numerical  simulations provide a 
fundamental instrument for investigating the phenomena occurring  within these systems.\\
The present paper is one such study, in which accretion disks with Advection Dominated Accretion Flows (ADAFs) 
\citep{narayan2012grmhd,abramowicz2013foundations, yuan2014hot}, the Magnetically Arrested Disk (MAD) 
\citep{narayan2003magnetically} and the Standard And Normal Evolution (SANE) \citep{narayan2012grmhd} disk 
configuration, are simulated using the equations of General Relativistic Magnetohydrodynamics (GRMHD) 
\citep{shibata2005magnetohydrodynamics,mizuno2024general}. 
SANE and MAD states represent limiting magnetization regimes of magnetized accretion flows in which the 
horizon-threading magnetic flux remains respectively subcritical or saturates at dynamically important 
values, respectively. The distinction is most clearly realized in geometrically thick, advective flows, 
where magnetic-flux advection is efficient, although analogous magnetically arrested states may arise 
in radiatively efficient disks provided sufficient net poloidal flux is supplied.\\ 
It is worth noting that the nomenclature ''Magnetically Arrested Disk'' can be somewhat misleading, 
as accretion is only ever temporarily suppressed and resumes after a transient time in a periodic 
fashion. While we use the SANE and MAD labels for convenience in the following, we also highlight which
fundamental properties of the flows or of the magnetic fields, including the dependence on BH spin, are 
at play in each instance, so that the results are not strictly tied to the specific type of simulation, 
but provide deeper insight into the underlying physics.\\

Once the accretion rate and angular momentum transport in our simulations reach a quasi-steady regime, energetic 
tracer particles are injected and evolved self-consistently with the disk in order to investigate their transport 
and propagation properties. 
These particles respond to the electromagnetic fields generated by the plasma flow, and obey the laws of 
General Relativity (GR) to account for strong gravitational effects. At the moment, these particles are completely 
passive and do not exert any feedback on their environment. In GRMHD simulations, the normalization length is set by the gravitational radius 
($r_g = GM/c^2 = 1.5 \times 10^5 (M/M_{\odot})$ cm, with $G$ the gravitational constant, $M$ the BH mass and $c$ the speed 
of light)
The ratio of the Larmor  radius ($r_L$, defined below) of the energetic particles to the gravitational radius, 
is what determines the scaling of the results of our simulations. Namely, the dynamics observed in these simulations 
are representative of any system in which particles move with GR dynamics and satisfy the ratio $r_L/r_G$ set by our parameters.\\
Overall, we aim to provide some further insight into the physics of particle acceleration and neutrino 
production in AGN environments. Indeed, reproducing the measured spectra of such particles requires modeling the emission by the 
distribution function at the source with certain assumptions (e.g., proton escape times in Fokker-Planck type 
diffusive modeling \citep[e.g.,][]{fraschetti2021effect, lemoine2025neutrinos}), that can be rendered more 
robust by supplying information from numerical simulations. At the same time, the simulations presented 
in this paper seek to provide a baseline for exploring the particle confinement properties of ADAFs, 
potentially identifying shortcomings and critical points that will need to be addressed  by future studies. 
For instance, classical ADAFs do not easily reproduce corona-like features, as the flows 
tend to be geometrically thick, with little temperature variations, except once the jet region 
\citep{nakamura2018parabolic,mizuno2022grmhd} is reached. For the purposes of this paper, 
the ``corona'' is defined as a region of gravitationally bound plasma near the black hole (within 
$\sim 30 \; r_g$) that satisfies the Bernoulli parameter criterion defined later and that is strongly
magnetized. This definition allows flexibility in identifying the corona's structure but explicitly 
excludes magnetized outflows (winds and jets) that are more common in ADAF simulations and that do 
not match observational constraints from NGC 1068 \citep{padovani2024high}. 
Thus, the analysis in this paper will help clarifying the potential of ADAFs as environments 
for energetic particle confinement, and their associated multi-messenger signatures. 
The paper is structured as follows: Sec.~\ref{sec_model} and Sec.~\ref{sec_setup} describe the numerical 
setup and simulation parameters. Sec.~\ref{sec_analysis} presents the analysis of particle escape and 
infall times, including comparisons with standard estimates, and examines the mechanisms of particle 
acceleration, with a focus on turbulent scattering. Sec.~\ref{sec_discussion} discusses the implications 
for high-energy astrophysics, including the potential role of these flows in accelerating cosmic rays 
and producing high-energy neutrinos. Finally, Sec.~\ref{sec_conclusion} summarizes the key findings and 
outlines directions for future work.

\section{Numerical methods} \label{sec_model}
The code employed for the simulations shown in this paper is a re-implementation of the Athena++ code 
\citep{stone2020athena++} based on the use of the KOKKOS library \citep{kokkos} for portability, named 
AthenaK \citep{stone2024athenak}. The advantages of having access to a multitude of architectures with 
good scalability, in particular GPUs, are manifest, and crucial when dealing with simulations that 
involve particles, since the computational cost can increase significantly (see App \ref{sec_performance}). 
The porting of codes to GPUs has also significantly contributed to making simulations 
of radiative flows more realistic \citep{skadowski2017radiative,jiang2019global,zhang2025radiation}. 
AthenaK uses Cartesian Kerr-Schild coordinates $(t,x,y,z)$ to define the lab frame in GR, which suits 
well the Cartesian grid employed by the code to discretize the physics.\\
The integration of the equations in GR is frequently done by employing a ``3+1'' decomposition 
\citep[e.g.,][]{thorne1986,rezzolla2013}, performing a so-called ``foliation'' of spacetime labeled
by coordinate time. This approach allows one to clearly distinguish temporal and spatial components 
of quantities measured by a ``fiducial'' (also called ``normal'') observer (FIDO) 
moving with 4-velocity (measured in the coordinate frame) $n_\mu^{\rm FD} = (-\alpha, 0,0,0)$ and 
$n_{\rm FD}^\mu = (1,-\beta^i)/\alpha$, with spacetime metric being expressed in the form
\begin{equation} \label{eq_adm_metric}
g_{\mu \nu} = \left[
\begin{array}{cc} - \alpha^2 + \beta_k \beta^k & \beta_i\\
\beta_j & \gamma_{i j}
\end{array} \right]\;,
\qquad 
g^{\mu \nu} = \left[ \begin{array}{cc}
- 1/\alpha^2 & \beta^i/\alpha^2\\
\beta^j/\alpha^2 & \gamma^{i j} - \beta^i \beta^j / \alpha^2
\end{array} \right]\;,
\end{equation}
where $\alpha = 1/\sqrt{-g^{00}}$ is the lapse function, $\beta^i = -g^{0i}/g^{00}$ is the shift three-vector, 
and $\gamma_{ij}$ is the spatial part of the spacetime metric $g_{\mu\nu}$, with $\gamma^{ij}$ 
the algebraic inverse of $\gamma_{ij}$. As usual, Greek indices run 
from $0$ to $3$, thus including ``temporal''  components, while Latin indices run from $1$ to $3$.\\ 
As we show below, the equations of motion for a charged particle can be incorporated as a Hamiltonian 
pair in terms of the shift vector $\beta^i$, the inverse of the spatial three-metric $\gamma^{ij}$ and 
the covariant particle four-velocity in coordinate basis $u_\mu = (u_0,u_i)$. 
Note that from 
Eq.~(\ref{eq_adm_metric}), the relation between $u_i$ and $u^i = dx^i/d\tau$ is $u_i = g_{i\nu}u^\nu 
= \beta_i u^0 +\gamma_{ij} u^j$. The FIDO measures the particle Lorentz factor and three-velocity to 
be $\Gamma = \sqrt{1+\gamma^{ij} u_i u_j}$ and 
$\tilde{v}^i = (v^i + \beta^i)/\alpha$, respectively, with ``coordinate'' (i.e., measured in a frame of 
reference at infinity) velocity $v^i \equiv dx^i/dt$. The FIDO and coordinate frame Lorentz factors 
are related as $u^0 = \Gamma/\alpha $, with $\Gamma$ the one measured in the FIDO frame.

\subsection{GRMHD in AthenaK}
AthenaK discretizes the fluid on a cartesian grid and applies Riemann solvers to solve the integral flux 
conservation equations, i.e., it uses a Finite Volume Method \citep{ripperda2019general,cheong2020gmunu,
stone2020athena++,mizuno2024general, stone2024athenak} and a Constrained Transport algorithm for the 
evolution of the magnetic field \citep{evans1988simulation}. For an in-depth analysis of the ideal GRMHD 
equations and their solutions, the reader is referred to the literature, as only some fundamental 
points will be recalled here.\\
The manifestly covariant equations can be summarized as follows: 
\begin{eqnarray}
&\partial_t (\sqrt{-g} \;  w^0 \rho) + \partial_j (\sqrt{-g} \;  w^j \rho) = 0\,, & \\
&\partial_t (\sqrt{-g} \;  T^0_\mu) + \partial_j (\sqrt{-g} \;  T^j_\mu \; \rho) 
= \frac{1}{2} \sqrt{-g} \;  \big( \partial_\mu g_{\lambda \kappa} \big) T^{\lambda \kappa}\,, & \\
&\partial_t (\sqrt{-g} \;  B^i) + \partial_j (\sqrt{-g} \;{}^*\!F^{ij} ) = 0\,. &
\end{eqnarray}
Here, $g:=\det(g_{\mu\nu})$ is the determinant of the spacetime four metric, with $\sqrt{-g} 
= \alpha\sqrt\gamma$, where $\gamma:=\det(\gamma_{ij})$ is the determinant of the spatial 
three metric, $w^\mu$ is the fluid four-velocity, $\rho$ is the proper rest mass density, 
$^*F^{\mu\nu}$ is the dual of the antisymmetric electromagnetic tensor, $T^{\mu\nu}$ is the 
energy-momentum tensor, and $B^i$ the Eulerian magnetic field in coordinate basis.
One can notice how the left-hand side of these equations obeys a flux conservation law, thus 
making the problem suitable for solution using Godunov-type schemes with Riemann solvers 
\citep[e.g.,][]{toro2009}. The magnetic field is evolved using a constrained transport scheme 
\citep{evans1988simulation} to ensure that the divergence-free constraint is enforced.\\
The GRMHD simulations are run by initializing a Fishbone-Moncrief 
\citep{fishbone1976relativistic} torus to which a dynamically negligible magnetic 
field is added in the form of poloidal loops encircling the torus's cross-section, cf.
eq.~(\ref{fishbone}).
The torus is initially in hydrostatic equilibrium, presenting an axisymmetric laminar flow with constant angular momentum, 
but the presence of the weak magnetic field allows the development of the Magneto-Rotational Instability (MRI) 
\citep{balbus1998instability} that enables accretion by providing a means for the 
outward transport of angular momentum. At the same time, MHD turbulence will develop as 
a consequence of the MRI, and such turbulence is bound to interact with energetic 
particles in the fluid, possibly leading to their stochastic acceleration 
\citep[e.g.,][]{lemoine2025neutrinos}, and/or to changes in the magnetic field configuration 
\citep{balbus1998instability}, that will affect the confinement of these particles. 
The self-consistent development of MRI-driven turbulence from an initial magnetic field 
provides the physical mechanism that has been otherwise modeled phenomenologically by the 
$\alpha$ parameter in \citet{shakura1973black, shakura1976theory}. 
The basic idea is always the same: stresses in the flow lead to accretion 
by effectively converting azimuthal motion into radial inflow, with the MRI offering 
a fundamental mechanism for this process that involves the Maxwell (magnetic) stress-tensor, 
whereas the classical model in  
\citet{shakura1973black, shakura1976theory} 
summarizes the effect of the Reynolds (fluid) stress-tensor with the parameter $\alpha$.
Depending on the amount of magnetic flux accreted at horizon scales and the mechanisms 
for its dissipation, the system will evolve into either the MAD or the SANE state, 
both of which are considered in this work. 
%
Ultimately, the initial conditions for the two systems differ only in the size of the initial 
torus. The magnetic field is kept dynamically negligible (hence the use of a hydrodynamic solution 
for the torus, with a weak, high-$\beta$ field added) and serves only to trigger the MRI and break 
the otherwise stable, constant angular momentum laminar flow, enabling accretion. The fully developed 
turbulent flow that follows exhibits notable differences from the initial conditions in both SANE 
and MAD regimes: magnetically dominated regions form, and in the MAD regime, accretion is modulated 
by the competition between gas and magnetic pressure. Accretion dynamics also self-consistently 
lead to the formation of outward flows, i.e., winds, that do not require strong magnetic fields to 
exist but can be enhanced and shaped by them, evolving into jet-like structures: highly collimated, 
strongly magnetized, relativistic outflows ejected along the poles of spinning black holes. These 
can feedback on scales far beyond those probed by the simulations. The fully developed system thus
exhibits a non-trivial structure, including vertical stratification, sheared and turbulent flows 
that connect spatially distant regions, and strongly varying magnetizations, among other key 
features.
While this general picture of the fully-developed flows is valid for both SANE and MAD regimes, 
the most notable, though certainly not the only, differences between the two are (i) the presence of 
bursty behaviour in MAD disks, with periodic ejections of matter that limit the build-up of magnetic 
pressure at horizon scales, (ii) the formation of jet-like structures in MAD disks 
\citep{blandford1977electromagnetic, 
livio1999extracting, narayan2012grmhd, toma2014electromotive, blandford2019relativistic, 
mizuno2022grmhd}, defined as regions of highly-magnetized outflows located at the 
poles of a BH, when the central BH has a sufficiently large spin parameter, and (iii) the development 
of non-axisymmetric modes in MAD regimes associated with the bursty behavior.
Note that SANE disks with high BH spin can also exhibit similar structures, but the
magnetization and stability (persistence over time) in MAD disks are higher, thus 
one refers broadly to such regions as ``funnel'' regions \citep{hawley2004general}. 
How these different accretion regimes affect energetic particle confinement is one
of the core questions addressed in this paper.
Although systems such as NGC 1068 do not seem to exhibit the dynamics and properties 
that match those of radiatively inefficient ADAFs \citep{narayan2012grmhd}, 
as these flows obey different radiative transport dynamics and operate at accretion 
rate well below $1 \%$ of the Eddington limit, ADAF models have proven effective for 
studying many features of supermassive black holes (SMBHs) \citep{porth2019event}. 
This suggests that the results of this study are relevant to real astrophysical 
scenarios, despite the GRMHD dynamics not fully matching those of any particular AGN 
one-to-one.\\
AthenaK uses normalized code quantities, including lengths, times, velocities, densities, 
and magnetic fields, to characteristic properties of the system. Physical values are 
recovered by specifying a mass, length, and density scale, so that, for example, 
velocities scale with $c$ and lengths with $GM/c^2$.

\subsection{Particles in AthenaK}
Particles in AthenaK are a recent development, and in order to investigate the physics 
relevant to this work, both GR and electromagnetic effects have to be accounted for in 
the particles' equations of motion. Since the space-time metric in our simulations is 
an analytical, static function of space only, it is possible to use the Hamiltonian 
formulation of the geodesic equation of motion in ``3+1'' decomposition to derive a 
conservative pusher for GR effects. Electromagnetic effects, on the other hand, are 
computed using discrete fields stored in the computational grid, and thus are not 
suitable for a Hamiltonian treatment, requiring other types of approaches. 
Following the work done in \citet{bacchini2018generalized, ripperda2018comparison, 
bacchini2019generalized, zou2024new}, we implemented an implicit mid-point scheme
with the Newton-Raphson method 
to compute the updated values of position and velocity.
The equations to solve are:
\begin{eqnarray} \label{eq_push_main}
& \frac{d}{dt}x^i = \alpha \,\frac{\gamma^{i k} u_k}{( \gamma^{j k} u_j u_k  
+ \epsilon )^{1/2}} - \beta^i\;, & \\
& \frac{d}{dt}u_i = - ( \gamma^{j k} u_j u_k + \epsilon )^{1/2} \partial_{x^i} \alpha 
- \frac{\alpha}{2} \frac{\partial_{x^i}(\gamma^{j k}) u_j u_k}{( \gamma^{j k} u_j u_k + \epsilon )^{1/2}} 
+ u_k \partial_{x^i} \beta^k + & \nonumber \\ 
&  + \frac{q}{m}\left( \alpha \, \gamma_{ij} D^j + e_{ijk} \frac{\gamma^{ih}u_hB^k}{u^0} \right)\;, \label{eq_push_main2} & 
\end{eqnarray}
where the geodesic part is derived from the free-particle Hamiltonian:
\begin{equation}
\Ham (x^i, u_i) = \alpha ( \gamma^{j k} u_j u_k + \epsilon )^{1/2} - \beta^j u_j\,,
\end{equation}
where $u_\mu$ is the covariant particle 4-velocity in the coordinate frame, and $\epsilon =1$ for massive 
particles. The electromagnetic terms are obtained by expressing the terms related to the Faraday tensor
$q/m \; F^{\mu \nu} u_\nu$ as a function of the quantities available in the code 
(fields in the coordinate frame and $3+1$ metric).\\
The relation for the electric displacement field $D^i$, as measured by a FIDO, is
\begin{equation}
\alpha D^i = E^i - \gamma^{ij} \: e_{jkl} \, \beta^k B^l\,,
\end{equation}
with Levi-Civita (pseudo)-tensor $e_{jkl} = \epsilon_{ijk}\sqrt{\gamma}$.\\
At each time-step, for each particle, non-linear iterations are performed to advance the 
values for the position and velocity at the next time-step until a tolerance of $10^{-16}$ on 
the (square of the) residuals is met. The six-dimensional system of 
equations~(\ref{eq_push_main})-(\ref{eq_push_main2}) is inverted by building the full 6-dimensional 
Jacobian matrix. While this is more expensive than a GMRES-like approach \citep{Saad1986} the 
straightforwardness of the implementation, the small size of the system and the small number of 
iterations typically needed to reach convergence (the average count is around $15$, with recorded 
maxima below $60$) do not warrant, at present, the employment of more sophisticated algorithms.\\
The particles' diagnostics consist of the position at each time-step of interest and the Lorentz 
factor measured by the normal observer ($\Gamma$).
Particles are removed from the domain over time as they cross either the outer boundary of the 
simulation or the inner boundary, set at $r = 2.5 \, r_g$ for reasons explained in Sec. \ref{sec_setup}. 
At present, particles are injected at a single specific time and no mechanism is implemented to 
have particles re-injected in the system, meaning that the total number of particles in the 
system decreases monotonically over time.\\
The details of the numerical scheme are available in the references, and salient points of 
the validation work done during development are reported in Appendix~\ref{sec_validation}.\\

\section{Simulation setup} \label{sec_setup}
\begin{figure}
\centering
\begin{subfigure}{\DefFigWidth}
    \includegraphics[width=\textwidth, trim = 0 0 0 8em, clip]{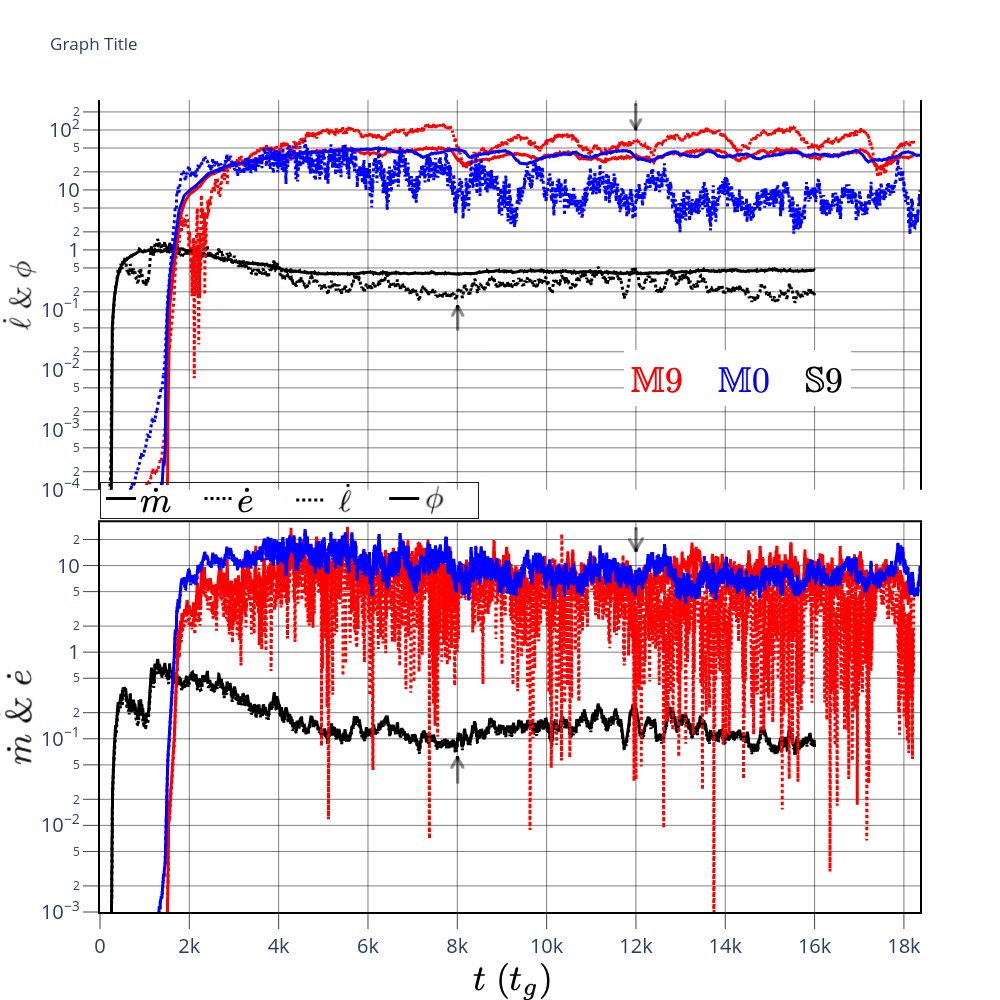}
    \caption{\label{fig_history} Comparison of the spherically integrated angular momentum transport 
    rate ($\dot{\ell}$), the magnetic flux ($\phi$), the mass accretion rate ($\dot{m}$) 
    and the energy transport rate ($\dot{e}$) in code units at $r = 12 \; r_g$. Black arrows indicate 
    the time of injection of the particles.}
\end{subfigure}\; 
\begin{subfigure}{\DefFigWidth}
    \includegraphics[width=\textwidth, trim = 0 0 0 8em, clip]{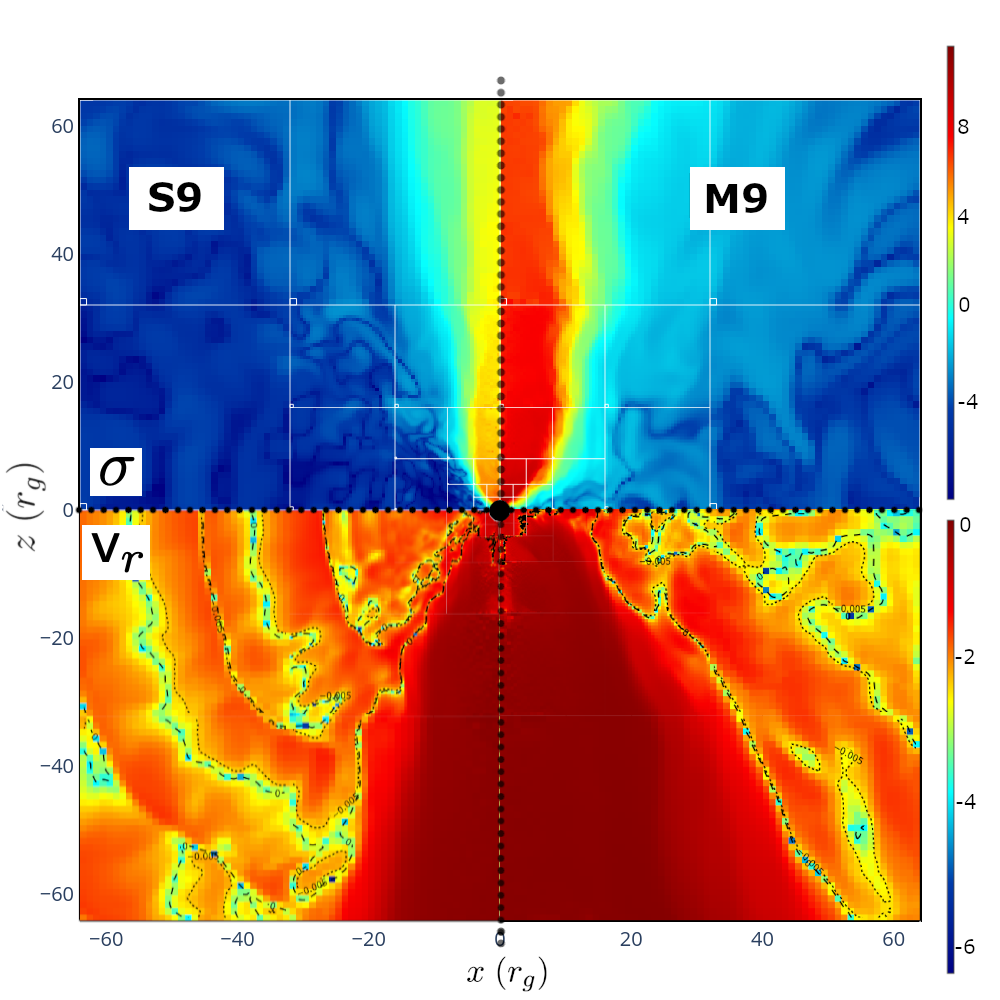}
    \caption{\label{fig_sigma} Slice at $y=0 \; r_g$, averaged over $50 \, t_g$ in the quasi-steady 
    phase, of the magnetizaton $\log(\sigma)$ (top half) and the radial (in the spherical sense) 
    3-velocity $\log(\mathsf{v}_r/c)$ (bottom half), comparing $\mathbb{M}9$ (right half) and 
    $\mathbb{S}9$ (left half).
    The dashed black line in the bottom half indicates the zero value where the direction of 
    $\mathsf{v}_r$ inverts, with an additional dotted line on the side of the negative values.}
\end{subfigure}
\caption{Summary of the most relevant global properties of the GRMHD simulations analyzed in this paper.}
\end{figure}

\noindent \textbf{Accretion disk}\\
The accretion tori are initialized as is usual for the SANE and MAD regimes, i.e., with parameters 
listed in Tab. \ref{table_torus_params}, where the main difference between the two initializations 
lies in the size and distance of the torus from the BH, and in the different dependence of the 
magnetic field on the hydrodynamic quantities. A SANE case with spin parameter $a=0$ was not 
included due to the excessively weak magnetic field for the parameters of interest. 
The Fishbone-Moncrief torus \citep{fishbone1976relativistic} is a hydrodynamic self-similar solution 
on top of which a weak (high plasma $\beta$) magnetic field is imposed. The field is defined through 
a magnetic potential $\psi$ that has only a azimuthal component, producing poloidal loops in the 
magnetic field, with the expression (in Boyer-Lindquist, BL, coordinates):
\begin{equation}\label{fishbone}
    \psi = \left( \frac{r}{r_{in}} \right)^{\alpha_\psi} \exp\left( -\frac{r}{\phi_\psi}\right) 
    \rho^{\omega_\psi}\;,
\end{equation}
where $r$ is the radius in BL coordinates, $\rho$ is the fluid density in the fluid rest frame, 
and $\alpha_\psi$, $\phi_\psi$ and $\omega_\psi$ are parameters used to shape the magnetic potential.
The exponential and the $\rho$-dependence only apply to the MAD case. 
The hydrodynamic Fishbone-Moncrief solution is characterized by two main parameters, $r_{in}$ and 
$r_{peak}$, 
that determine the distance of the innermost edge of the torus from the central BH and the radial 
position of the maximum of the enthalpy $h$ (defined in Sec. \ref{sec_acceleration}), respectively.\\
The resolution of the grid is set to $128 \times 128 \times 128$ in the region (in units of $r_g$) 
$[-4,4]\times[-4,4]\times[-4,4]$, with Static Mesh Refinement (SMR) that halves the grid resolution 
each time the domain extent is doubled, with the total box size set to $[-512,512] \times [-512,512] 
\times[-512,512]$.\\
The GRMHD parts of the MAD (SANE) simulations are allowed to evolve for $t = 1.2 \times 10^4 \; 
t_g = 1.2 \times 10^4 \; r_g/c$ ($t = 10^4 \; r_g/c$), until a quasi-steady regime in the accretion rate, 
angular momentum and magnetic flux transport is reached (see Fig. \ref{fig_history}). For the resolution 
chosen (i.e., $r_g/16$ in the innermost region), it has been shown \citep{porth2019event} that 
the macro-scale dynamics of the disk have saturated, in the sense that increasing the resolution
further does not lead to significant changes in the accretion rates or the magnetic fluxes. While 
increasing the resolution would allow the study of particles with lower energies, the limited gain 
in the dynamic range of particle energies does not justify the substantial rise in computational 
cost at this exploratory stage. Moreover, maintaining a lower resolution enables running multiple 
simulations to explore the parameter space more effectively.\\
The choice of resolution is also informed by the properties of the MRI, as it can be shown 
\citep{balbus1998instability}, that for a given rotation profile (assumed Keplerian for simplicity), 
the maximum growth rate of the MRI in the disk is given by 
$\gamma_{\max} = - (R/2) \; d \Omega/d R \approx (3/4)\,c\,\sqrt{r_g/R^3} \leq \gamma_{\max} 
(R=8 \, r_g) \approx 0.03  \, t^{-1}_g$ 
(the reference position $R=8 \, r_g$ being chosen to lie within the disk of the 
SANE cases, that are much smaller than the MAD cases, and thus have higher growth rates),  
with wavenumber 
$k_{\max} \approx \Omega/v_A\;\sqrt{15}/4 \approx \sqrt{15/32} \;
\sqrt{\beta} \;\Omega/c_s \approx \sqrt{15/32} \; \sqrt{\beta} \; (R/H) (1/R) \approx 0.427 
\; r^{-1}_g$ (for the innermost boundary of the jet, where $\beta \approx 100$, $R/H \approx 1/2$)
which is well resolved in the present simulations of high-$\beta$, thick tori ($H\sim R$).\\

\noindent \textbf{Injection of tracer particles}\\
Particles are injected at $t=8 \times 10^3 \; t_g$ in the SANE case and at $t=1.2 \times 10^4 \; t_g$ 
in the MAD case, within a spherical shell at $r \in [7.5,8.5] r_g$, with velocities aligned with 
the local fluid flow.
Initial Lorentz factors $u^0_{\rm init}$ (see Tab.~\ref{table_torus_params}) are drawn from a 
uniform distribution, and particles are assigned a realistic charge-to-mass ratio of $e/m=10^8$ 
[in SI units: C/kg] (for protons).
Because of the realistic $e/m$ ratio chosen, the gyration frequencies of particles can quickly 
grow to values that impose excessively small time steps to resolve the gyration. 
Focusing on particles with higher $\Gamma$ factors increases the Larmor frequency, rendering the 
computational task manageable and provides an interesting scenario to investigate the physics of 
cosmic-ray acceleration.\\
The ranges of $\Gamma$ factors used to initialize particles are chosen such that, at their (spherical) 
injection location $r = 8 \; r_g$, their Larmor radii satisfy $5 \leq r_L/r_g = \Gamma m v^\perp/
(\alpha e B \; r_g) \vert_{r = 8 \,r_g} \leq 25$, assuming initial velocities purely perpendicular to 
the field. Here, $r_L$ is measured in the coordinate frame, in which $B^i$ is defined. This choice 
ensures that the Larmor orbits are resolved well for the spatial resolution of the fluid grid, even 
as particles enter the innermost regions where the magnetic field increases significantly. The broad 
range of Larmor radii further reflects the intention of maintaining a consistent energy distribution
across all injection locations. Incidentally, since both the magnitude of the metric coefficients and 
of the magnetic field magnitude $B$ scale approximately as $1/r$, in the absence of dynamics other than 
gravity, one only needs to ensure resolution of the orbits in the innermost regions with largest $B$ and 
will automatically be resolving the Larmor motion at all position. For realistic applications, where there 
are additional dynamics, a margin of safety should be considered, and additional points in regions 
where the magnetic field is strong but the resolution weaker than closest to the BH should be 
monitored for under-resolution, as done here.\\
Indeed, with these initialization parameters, assuming constant, or slightly decreasing, particle 
energy during motion, the spatial resolution of orbits is ensured, for the high-spin MAD case, up 
until $r = 2.5 \; r_g$. 
The high-spin MAD case is used as reference since it is the one with highest magnetic field closer 
to the event horizon. This requires that the artificial particle sink be set at this radius, and for 
ease of comparison we set the sink at the same radius across all simulations. 
At this radius, for the parameters chosen, particles have Larmor radii satisfying $0.5 \leq r_L/r_g 
\leq 2.5$. Future studies will aim at using finer resolutions for the GRMHD grid, that will allow to move 
the artificial sink closer to the actual event horizon, and thus probe dynamics even closer to the BH.\\
For the simulations presented in this paper, a resolution of $1 \; r_g$ or better (the inner most region 
has a resolution of $1/16 \; r_g$) is maintained until the domain reaches $[-64, 64] \times[-64, 64] 
\times[-64, 64]\;r_g $. Crucially, as one moves further away from the BH, the magnetic field drops 
in magnitude, leading to larger orbits and compensating for the weaker resolution. Notably, in the SANE 
simulation, the fall-off of the magnetic field is quicker than in the MAD cases, thus interesting dynamics 
in this simulation end up being mostly focused in the innermost region, with particles quickly switching 
to an almost free-streaming regime.\\
The chosen Larmor radius range for initialization also sets the scaling for the simulation results, as 
the particle dynamics scales with the ratio $r_L/r_g$. While the Larmor radius must be resolved by 
the grid and the Larmor frequency by the time step, the Larmor radius must also be large enough to allow 
interaction with turbulent structures in the flow, as these are the ones that can impact the propagation 
properties. This is why the ratio $5 \leq r_L/r_g \vert_{r = 8 \,r_g} \leq 25$ was chosen for this 
study, as it provides a balance of resolution and interesting dynamics. 
Simulations where the initialization was chosen to satisfy the $r_L/r_g$ criterion directly (rather 
than having energies within a certain range) were also run, with no remarkable differences from those 
presented here. However, setting the Larmor radius rather than the energy leads to a bias where the most 
energetic particles are located preferentially at regions of higher magnetic field, which is not 
desirable for obtaining a comprehensive picture of the particle dynamics. 
The Larmor frequency of the injected particles $\omega_L = \alpha eB/( 2\pi \Gamma m) \; t_g^{-1}$ 
has to be resolved, thus we implement an adaptive time-step solution that ensures the condition 
$\Delta t < 0.5 / \omega_L$ is satisfied, from the time-point when particles are injected onwards, 
i.e., the Larmor orbits are resolved with at least two points for all particles (the worst-case 
scenario of two points likely occurring for only a few particles in the most extreme regions). 
Since AthenaK sets a global time-step, this then becomes the time-step for the evolution of the 
system \citep{stone2024athenak}.
Since the pusher we are using is implicit, its stability is guaranteed, regardless of how many points 
are used to resolve the orbits. However, in order to sample the interaction of particles with the 
fluid, the more points one can use to resolve the orbit, the better. 
As particles leave their injection locations towards regions of higher magnetic field, the time step 
can decrease by up to two orders of magnitude. The chaotic nature of the particle dynamics also 
introduces unpredictability in the evolution of the adaptive time step; jumps of more than one order 
of magnitude between cycles can, for example, be observed when a particle that was extremely close 
to the horizon ends up being swallowed.\\ 
For reference, the time step used to evolve the GRMHD simulations alone is $\Delta t = 1.25 \times 
10^{-2}\;t_g$, the injection of particles leads to an average time step $6.0 \times 10^{-4} \; t_g 
\leq \Delta t \leq 3.6 \times 10^{-3} \; t_g $, but temporary drops to $\Delta t \approx 4.0 \times 
10^{-4} \; t_g $ are observed.
Particles and GRMHD fluid are evolved together for an additional $\Delta t = 2 \times 10^3 t_g$ (a 
little over one orbital time at the original position of the pressure peak in the MAD tori) in the 
MAD simulations and $\Delta t = 10^3 \; t_g $ for the SANE cases, after which time most particles 
have left the innermost region of interest to our study.\\
Injecting particles with velocities exceeding those of the background flow means that an injection 
mechanism capable of generating particles of such energies is assumed, but not directly simulated, 
a limitation of the current approach that will need to be addressed in future work. On the other 
hand, it is highly interesting to examine the dynamics of energetic particles, as it is not trivial 
to determine their fate solely on heuristic scale arguments (e.g., the Hillas criterion 
\citep{hillas1984origin}). In general, injected particles should not have excessively large Larmor 
radii, as the magnetic field generally satisfies (outside the jet region and some very localized 
areas) $\beta \geq 1$ and thus provides little contribution in confining particles that have energies 
(scaled to the Larmor radius) significantly exceeding that of the bulk fluid flow. Intuitively, the 
initial torus is in hydrostatic equilibrium, in a stable orbit with Keplerian velocity. If its velocity 
were any larger, it would escape from its orbit, and likewise particles with energies significantly 
larger will be almost free to escape the system without feeling any confinement effect from the disc. 
Furthermore, analytical approaches, tested through simulations, to the interaction of particles and 
MHD-like turbulence \citep{lemoine2023particle} suggests that scales much below the Larmor radius will 
not have significant effects on the transport of particles.\\
A total of $N_p \approx 1.5 \times 10^6$ particles is injected in each simulation. A summary of 
the simulation runs and their parameters is shown in Table \ref{table_torus_params}.

\begin{table}
\begin{ruledtabular}
\begin{tabular}{|cccc|}
\textbf{Parameter} & \textbf{MAD} ($\mathbb{M}0$) & \textbf{MAD} ($\mathbb{M}9$) & \textbf{SANE} ($\mathbb{S}9$)\\ \hline
$r_{in}$ & 20.0 & 20.0 & 6.0 \\
$r_{peak}$ & 41.0 & 41.0 & 12.0 \\
$\beta_{\min}$ & 100.0 & 100.0 & 100.0 \\
$\alpha_\psi$ & 3.0 & 3.0 & 1.0 \\
$\phi_\psi$ & 400.0 & 400.0 & -- \\
$\omega_\psi$ & 3.0 & 3.0 & -- \\
$a$ & 0 & 0.9375 & 0.9375\\
$u^0_{\rm init} \in$ & [$5.0\cdot 10^7$, $2.5\cdot 10^8$] & [$5.0\cdot 10^7$, $2.5\cdot 10^8$] & [$10^7$, $5.0\cdot 10^7$]\\
\end{tabular}
\caption{\label{table_torus_params} Summary of the torus (eq.~\ref{fishbone}) and particle injection parameters 
used in the GRMHD simulations. Energetic particles are injected in a spherical shell with $r \in [7.5,8.5] r_g$.
A zero-spin BH ($a=0$) is explored only for the MAD case.}
 \end{ruledtabular}
\end{table}

\section{Analysis of results} \label{sec_analysis}
We first perform a statistical analysis of the particle dynamics to identify general trends and guide the subsequent 
analysis, that consists of an in-depth look at the histories of individual particles to understand how the interaction 
of the particles with the MHD accretion flow determines their dynamics.\\
Different particle initializations were tested (e.g., initialization based on a density threshold in the fluid or at 
different radii). These did not lead to significant changes in the properties reported here, but made it more difficult 
to assess quantities like the escape times, as it becomes less straightforward to compare escape times for particles 
initialized at the random location where the fluid meets certain criteria. Furthermore, not having control over where 
particles are initialized makes it more difficult to identify a set of parameters that ensure proper resolution of 
the orbits for all particles involved in the study.

\subsection{Statistical Analysis}
In order to carry out the statistical analysis of the results, a new, spatial (cylindrical) grid, independent
of the one used to carry out the simulations, is introduced during post-processing. 
Such a grid is used to ensure that enough particles are being sampled when estimating statistical quantities, 
and to facilitate the mapping of the particle quantities to the fluid dynamics, as the fluid quantities can 
also be mapped to this new grid by coarsening or interpolation, depending on the relative resolution of the 
simulation and post-processing grid. 
The grid is also discretized in energy space for the particles. The spatial domain of the simulation within 
a box $[-64, 64] \times[-64, 64] \times[-64, 64]\;r_g $ is projected to 
cylindrical coordinates $(R,\varphi,z)$ with a resolution $(30,20,30)$. 
Initially, 25 linearly spaced bins in the energy domain $\Gamma \in [10^7, 3 \times 10^8]$ (MAD) and 
$[2 \times 10^6, 5 \times 10^7]$ (SANE) are used to study the overall particle behavior. To capture 
acceleration to very high energies and avoid artificial accumulation near the injection upper limit, 
a broader analysis with 25 logarithmically spaced bins over $\Gamma \in [10^6, 5 \times 10^{10}]$ is 
also performed (see Fig.~\ref{fig_avg_dist}), enabling the observation of a power-law regime in the 
distribution.\\
Additional analysis is carried out on a grid with resolution $(100,20,100)$, covering the domain $[-500, 500] 
\times[-500, 500] \times[-500, 500]\;r_g $, in order to examine the dynamics further away from the BH.  \\
Figures use colormaps to represent the quantity that is the subject of the figure, which is, for the smaller 
$[-64, 64] \times[-64, 64] \times[-64, 64]\;r_g $ domain, an average involving hundreds to thousands of particles 
for ``single-contribution'' quantities (e.g., the escape time, since each particle can contribute only once to 
only one cell) and thousands to tens of thousands for ``multiple-contribution'' quantities (e.g., the trapping 
time or the $\Delta \Gamma$, as each particle can contribute at multiple locations with different energies at 
different times). Having a sufficiently large number of particles contributing to the averaging is particularly 
important for energy-related quantities, since particle gyration induces rapid, small-amplitude oscillations 
in the Lorentz factor (see Fig. \ref{fig_inhst_det}). These oscillations can be neglected only when averaging 
over enough particles for their net contribution to cancel out.
To facilitate interpretation of the results, contour levels of the magnetization $\sigma = b^\mu b_\mu/(\rho c^2) 
= 1$ (where $b^\mu$ is the magnetic field in the fluid rest frame, obtained from the Hodge dual of 
the electromagnetic tensor $ b^\mu = {}^* F^{\mu \nu} w_\nu $) and the rest-frame density $\rho = 0.001$, 
determined from the fluid diagnostics, are also shown (based on data averaged over $\Delta t > 300 \; t_g$).
These values provide a reference to identify where the magnetic energy becomes dominant over the fluid rest 
mass, as well as separating the disk from the low-density regions.\\
The analysis of the particles' trajectories is done with a temporal resolution of $\Delta t 
= 6 \; t_g$ (diagnostics are saved every $2 \; t_g$).\\

\subsubsection{Escape Times} \label{sec_escape}
\noindent \textbf{Estimating escape times from numerical simulations}\\
Having access to the full history of the particles in the system allows for the study of many 
characteristics of their dynamics. In this section, the focus will be on the escape and infall times 
($\tau_{esc}$ and $\tau_{in}$ respectively) of the particles injected as described in Sec. \ref{sec_setup}. 
Here, $\tau_{esc}$ is defined as the difference between the last time-point before a particle leaves the region of 
interest at the outer boundary (i.e., at $R \leq 64 \; r_g$ and $|z| \leq 64 \; r_g$) and the time the particle 
was injected. Similarly, $\tau_{in}$ is defined as the difference between the last-time point before 
a particle leaves the domain at the artificial sink at $r = 2.5 \; r_g$ and the time it was injected. 
This allows to reconstruct maps that indicate the distribution of $\tau_{esc}$ and $\tau_{in}$ within the disk, 
that can be averaged over the energy of the particles or further broken down by energy range. 
$\tau_{esc}$ does not take into account particles that might re-enter the region of interest 
at a later time (see Sec. \ref{sec_individual}), and is thus more applicable to the usual approach 
(e.g., \citep{fraschetti2021effect,lemoine2025neutrinos}) of assuming particles are injected at one 
location and eventually diffuse out of the region of interest after a typical time $\tau_{esc}$.\\
The choice of the outermost point at $R = 64 \; r_g$ and $|z| = 64 \; r_g$ for the more detailed 
analysis carried out here is somewhat arbitrary, but it can be justified. 
The dynamics that are specific to a GR environment, such as the generation of 
powerful outflows \citep{blandford1977electromagnetic, blandford1982hydromagnetic}, 
and that are likely to have the strongest effect on particles, occur in the innermost parts of the accretion flow (see 
Fig.~\ref{fig_maps_combined}), as this is where the metric deviates the most from the flat-space 
metric and the dynamics of the flow are most affected by non-Newtonian effects. Secondly, with the long-term goal 
of applying these results to interpret the neutrino detection by IceCube \citep{icecube2022evidence}, 
the region of the accretion flow that is of greatest interest for this purpose is the so-called 
``corona'' \citep{liang1979hard, murase2022hidden, padovani2024high, 
middei2026directly, dash2026accretion, dhaka2026accretion}, 
that is inferred to have an extension of a few tens of $r_g$. 
Lastly, as is usual for GRMHD simulations, the initial condition of the weakly magnetized torus should 
be sufficiently evolved to have dynamics that are not strongly dependent on the initial setup 
\citep[e.g.,][]{narayan2012grmhd}.  
In the outer parts of the torus, in particular along the mid-plane where the outflows from the BH 
do not have direct effects, this requires much longer evolution times than those considered in these 
simulations, and thus some artifacts are likely to appear (see Sec. \ref{sec_individual}).\\
There is also a technical reason, as mentioned in Sec. \ref{sec_setup}, since the resolution 
used in these simulations only allows to resolve the dynamics of particles with $r_L/r_g \approx 1$ 
up to $R \leq 64 \; r_g$ and $|z| \leq 64 \; r_g$, thus dynamics further out are less trustworthy, 
even though the drop in magnetic field means that Larmor radii can grow significantly and it 
will be shown (see Sec. \ref{sec_individual}) that interesting dynamics can occur even further 
from the BH. 
Note that at sufficient large distances the lapse function $\alpha \approx 1$, meaning that the 
Lorentz factor measured at infinity is captured by the quantity $\Gamma =  u^0 \alpha$, which will 
be convenient when analyzing the particle distributions in Sec.~\ref{sec_acceleration}. We often 
use the Lorentz factor $\Gamma$ as a proxy for the energy of a particle at infinity, $E_\infty = -u_0$, 
with the latter given by $u_0 = (-\alpha^2 + \beta_k \beta^k) u^0 + \beta_i u^i =-\alpha\Gamma +\beta^i u_i$.
Far enough from the BH $\alpha = 1$ and $\beta^i = 0$, whereas closer in, 
the presence of a gravitational well requires care in distinguishing changes 
in the total energy from changes in the kinetic energy (the latter captured by $\Gamma$). \\
Table \ref{table_escape} summarizes the main results regarding the escape times for all simulations  
broken down by the spawning energy of the particles. In SANE disks, weaker magnetic fields that fall-off
more rapidly with distance from the BH lead to larger gyro-radii, limiting the interesting dynamics 
to low-energy particles. 
Overall, of the particles initialized within the spherical shell, those whose poloidal coordinate 
positions them in the funnel-sheath regions (i.e., in regions where $\sigma \approx 1$) 
have the largest $\tau_{esc}$, which may seem surprising: in these simulations, particles are initialized 
with velocities aligned with that of the fluid, that close to the BH, and away from the mid-plane, is 
directed as a strong outflowing wind. One could thus expect particles in this region to leave the system 
quickly. However, the funnel-sheath region is where the strongest turbulence is located, and, as will be 
shown, turbulent scattering plays a major role in the dynamics of the particles, including several 
deviations of the particles' trajectory, that increase the time they remains in the region.\\ 
Table \ref{table_escape} shows, somewhat unsurprisingly, that the highest energy particles mostly escape 
at infinity in these simulations. This is readily explained by the fact that the particles in question 
have large Larmor radii, and are thus less confined by the magnetic field. Other than the magnetic field, 
the only confining factor is the gravitational potential, that has weak effects far from the BH horizon 
on particles at such high energies.\\
Nevertheless, a significant fraction of particles are captured by the artificial sink at $r = 2.5 \; r_g$
(see Table \ref{table_escape}), even though their energies may lie in the higher end of the spectrum. 
This may reflect the acceleration of lower-energy particles through interactions with the disk as they 
approach the innermost regions of the accretion flow.\\

\noindent \textbf{Comparing simulation results to analytical estimates}\\
To explore the overall transport behavior of particles in the system, their statistically averaged 
properties are compared to simple estimates from three canonical transport models: advection, 
free-streaming, and uniform diffusion. If the particle dynamics match the advection estimate, 
then particles effectively behave as "frozen-in" to the fluid, with their transport governed by 
fluid motion. In contrast, free-streaming represents the opposite extreme, with particles moving 
independently of the fluid, responding only to gravity. Since the GRMHD simulations show that 
the poloidal velocity $\mathsf{v}_{\theta}$ is, on average, one order of magnitude smaller than 
the azimuthal velocity $\mathsf{v}_{\varphi}$ (and thus of $\mathsf{v}_r$, see Tab. \ref{table_estimates}), 
our free-streaming estimate assumes that a particle has only azimuthal and radial velocity at 
injection, such that the ratio between the two is sufficient to characterize the velocity considering 
that $\mathsf{w}^i \mathsf{w}_i = c^2$. The diffusive regime, often modeled via Fokker-Planck 
approaches, can incorporate varying levels of complexity \citep[e.g.,][]{effenberger2018relation,
fraschetti2021effect,lemoine2023particle,lemoine2025neutrinos}, with the simplest case assuming 
constant (isotropic) diffusion in space and momentum. In the present work, we adopt an idealized 
diffusion coefficient as given by
\begin{equation} \label{eq_diff_coeff}
    \nu = \frac{(\Delta R)^2 + (\Delta z)^2}{2 \; \eta \; \tau_{\rm trap }}\,,
\end{equation}
with values for $\tau_{\rm trap }$ provided in Tab. \ref{table_trapping}, and the geometrical 
factor $\eta = 4$ for diffusion in 2D.\\
Since the particles in our simulations have large Larmor radii, a well-defined diffusive behavior 
may, however, not be reached. A uniform diffusion approach with the nominal coefficient above, 
in fact, breaks down quickly when comparing the average measured particle properties, summarized 
in Tab. \ref{table_escape}, to the estimates in the tenth column of Tab. \ref{table_estimates}. 
While a more sophisticated choice for the diffusion coefficient might have given more satisfactory 
comparisons
the significant gaps between the values in Tab. \ref{table_escape} suggest that uniform diffusion 
alone may not sufficiently capture the transport dynamics. The exploration of alternative models, 
such as space- and energy-dependent or anisotropic diffusion, is however left for future work.\\
The high-spin MAD case $\mathbb{M}9$ has infall times $\tau_{\rm in}$ that can be significantly 
larger than those determined from advection, depending on the poloidal region considered, and 
exceed the free-streaming estimate by more than a factor of ten.
Because of the smallness of $\mathsf{v}_{\theta}/\mathsf{v}_{\varphi}$ (and thus $\mathsf{v}_{\theta}/
\mathsf{v}_{r}$), the area considered for diffusive transport of particles is a circular sector 
covering $\Delta\theta = 0.5$.
The escape times for particles in the higher energy range (see Tab. \ref{table_escape}) can 
approach free-streaming estimates, indicating that particles with large Larmor radii interact 
weakly with the magnetic field in the accretion disk.\\ 
This behavior is also observed in $\mathbb{M}0$. $\mathbb{M}0$ also shows estimates for infall times from advection that are 
compatible with values measured directly from simulations, especially further away from the 
midplane, though advection severely overestimates the escape times. This suggests that if
particles do not spend sufficient time interacting with the disk, advection may provide a
reasonable estimate of their motion.
Similarly, the (uniform) diffusive estimate is reasonable for $\tau_{in}$ in $\mathbb{M}0$, 
primarily because particles, initialized with velocity aligned with the fluid, have the 
largest (relative) radial velocity in this case, making poloidal diffusion a more accurate 
proxy than in other configurations.\\ 
In $\mathbb{S}9$, escape times are compatibles with free-streaming estimates, as expected from
the distrint magnetic field scaling in MAD and SANE regimes (Sec.~\ref{sec_setup}). Particles
in $\mathbb{S}9$ interact significantly with the magnetic field only at radii smaller than 
their injection radius, but are almost freely streaming further out, as their Larmor radii 
become too large for interaction. This is supported by infall times, that are significantly 
longer than the free-streaming estimate, while also not matching the advection estimate, 
indicating more complex interactions between particles and fluid close to the BH. 
Although escape times are compatible with free-streaming, they are systematically smaller,
suggesting that interaction between particles and fluid close to the BH tends to deflect 
particles radially outwards, giving them additional radial velocity. This effect might be 
aided by gravitational assists from the BH, though particles in the simulations rarely 
complete undisturbed orbits due to frequent scattering events.\\  
Overall, $\mathbb{M}9$ provides the clearest example of particle dynamics deviating from
simple estimates based on fluid properties or uniform diffusive models.\\
There seems to be a ``trend'' (not shown) across the simulations where particles initialized at the lowest 
energies  tend to escape along the jet (or, more broadly, the central region of higher magnetization, 
i.e., the funnel), whereas particles with higher energies escape more isotropically, with a slightly 
over-density around the mid-plane. The relevance of this result is, however uncertain, as not enough 
realizations of the simulations have been carried out to assess whether this is an effect related to the 
particular initial conditions chosen.\\ 
Overall, Table \ref{table_escape} points to $\mathbb{M}9$ (high spin) as being the simulation where 
particles stay longest in the system at the lower end of the energy spectrum. 
This is in line with the magnetic field being the dominant driver for the dynamics of the 
particles as explored in the following. In $\mathbb{M}9$, the magnetic field magnitude decreases more 
slowly with radius \citep{pathak2025simulating} due to the bursty behavior that effectively ejects 
magnetic field into the disk with magnitude obtained by expanding the flux that was ``frozen-in'' at h
orizon scales. This means that, for the energy range chosen, interaction between the magnetic field 
and the particles extends for larger regions of the domain compared to the other simulations.\\
The parameter range chosen for $\mathbb{M}9$ seems to identify a threshold in the particle energy 
at $\Gamma \geq 2.0 \cdot 10^8$ where particles initialized in the disk region (with fluid density 
$\rho \geq 10^{-3}$) have a significantly decreased likelihood of falling into the BH. This is probably 
a combined effect of the higher energy and the more effective scattering against the turbulence in 
the region for particles with a particular $r_L/r_g$, but more detailed studies than those 
presented here are needed to draw firm conclusions. It will be particularly interesting to extend 
the energy range of injection to potentially identify other similar thresholds, also for the other 
cases, as well as regarding escape.\\
Expanding the region of interest for this analysis to $R \leq 500 \; r_g$ and $|z| \leq 500 \; r_g$, 
the average $\langle \tau_{esc} \rangle$ exceeds $1.2 \cdot 10^3\; t_g$, in the MAD cases, which is 
still larger than the naive expectation $\langle \tau_{esc} \rangle \leq 500 \cdot \sqrt{3} \approx 
870 \; r_g / c$, and shows less variance with respect to particle energy. Infall times
$\langle \tau_{in}\rangle$ can also increase, but this effect is only significant in $\mathbb{M}9$, 
where a dramatically increased $\langle \tau_{in}\rangle \geq 10^3 \; t_g$ is recorded. 
This is related to the much stronger effect of turbulence on particles in $\mathbb{M}9$, and 
to the phenomenon analyzed in Sec. \ref{sec_individual} of particles being reflected 
multiple times while traveling along the funnel, which makes it more likely for them to end up 
inside the BH.\\
These results highlight the necessity of stochastic models that capture both coherent transport and 
turbulent scattering to accurately describe particle dynamics in magnetized accretion flows.

\begin{table}
\begin{ruledtabular}
\begin{tabular}{|cccccc|}
\textbf{Simulation} & $\Gamma $ & $<\tau_{in}> $ & $<\tau_{esc}> $ & $N^{\Gamma}_{in}/N^{tot}_{in}$ & $N^{\Gamma}_{esc} / N^{tot}_{esc} $\\ \hline
\multirow{2}{*}{$\mathbb{M}9$} & $[0.1, 1.5] \times 10^8$ & $[78, 280] \; t_g$ & $[72, 302] \; t_g$ & $ 0.87 \; (81644/94138) $ & $ 0.43 \; (267956/618493) $ \\
& $[2.0, 3.0] \times 10^8$ & $[73, 222] \; t_g$ & $[66, 90] \; t_g$ & $0.05 \; (4243 / 94138)$ & $0.25 \; (154996 / 618493)$ \\ \hline
\multirow{2}{*}{$\mathbb{M}0$} & $[0.1, 1.5] \times 10^8$ & $[8, 70] \; t_g$ & $[90, 180] \; t_g$ & $ 0.61 \; (157011 /258111 )$ & $ 0.40 \; (242827/612119)$\\
& $[2.0, 3.0] \times 10^8$ & $[10, 70] \; t_g$ & $[72, 114] \; t_g$ & $ 0.16 \; (40211/258111)$ & $ 0.26 \; (158271/612119)$\\ \hline
\multirow{2}{*}{$\mathbb{S}9$} & $[0.2, 1.5] \times 10^7$ & $[70, 130] \; t_g$ & $[78, 100] \; t_g$ & $ 0.14 \; (5018/34773)$ & $0.07 \; (80363/1091104)$ \\
& $[4,0, 5.0] \times 10^7$ & $[70, 72] \; t_g$ & $[77, 92] \; t_g$ & $ 0.21 \; (7398/34773)$ & $0.24 \; (258854/1091104)$
\end{tabular}
\caption{\label{table_escape} Summary of different properties for particles in different energy regimes related 
to the analysis within the region $R \leq 64 \; r_g$ and $|z| \leq 64 \; r_g$. $\tau_{esc}$ refers to the time 
it takes particles to leave this domain at the outer boundaries after being spawned. $\tau_{in}$ refers to the 
time it takes particles to fall inside the BH region after being spawned. $N^{\Gamma}_{esc}/N^{tot}_{esc}$ and 
$N^{\Gamma}_{in}/N^{tot}_{in}$ refer to the fraction of particles that either escape at the outer boundary or 
fall into the black hole region at a given energy range $\Gamma$ with respect to the total number of particles 
that have (respectively) escaped and fallen in.}
\end{ruledtabular}
\end{table}
 
\begin{table}
\begin{ruledtabular}
\begin{tabular}{|cc|c|cc|ccc|cc|}
\textbf{Simulation} & in/esc & $\theta$ & $\mathsf{v}_r / c$ & $\tau_{\rm adv} / t_g $ & $\mathsf{v}_{r}/\mathsf{v}_{\varphi}$ 
& $\mathsf{w}_{r} / c$ & $\tau_{\rm fs} / t_g$ & $\nu / (r_g \cdot c)$ & $\tau_{\nu} / t_g$\\ \hline
\multirow{6}{*}{$\mathbb{M}9$} & \multirow{3}{*}{in} & $0.0$ & $0.05$ & $110$ & $2.74$ & $0.94$ & $6$ & \multirow{3}{*}{$0.255$} & \multirow{3}{*}{$28$} \\
 & & $0.349$ & $0.103$ & $53$ & $3.47$ & $0.96$ & $6$ & & \\
 & & $\pi/4$ & $0.378$ & $14$ & $18.21$ & $0.99$ & $5$ & & \\ 
 \arrayrulecolor{lightgray} \cmidrule{2-10} \arrayrulecolor{black}
& \multirow{3}{*}{esc} &  $0.0$ & $0.05$ & $1.12 \cdot 10^3$ & $1.2$ & $0.77$ & $72$ & \multirow{3}{*}{$0.318$} & \multirow{3}{*}{$2.45 \cdot 10^3$} \\
 & & $0.349$ & $0.103$ & $544$ & $3.47$ & $0.96$ & $58$ & & \\
 & & $\pi/4$ & $0.378$ & $148$ & $18.21$ & $0.99$ & $56$ & & \\ \hline
\multirow{6}{*}{$\mathbb{M}0$} & \multirow{3}{*}{in} & $0.0$ & $0.05$ & $110$ & $1.56$ & $0.84$ & $6$ & \multirow{3}{*}{$0.37$} & \multirow{3}{*}{$20$} \\
 & & $0.349$ & $0.14$ & $39$ & $5.89$ & $0.98$ & $5$ & & \\
 & & $\pi/4$ & $0.272$ & $20$ & $92.74$ & $0.99$ & $5$ & & \\
 \arrayrulecolor{lightgray} \cmidrule{2-10} \arrayrulecolor{black}
& \multirow{3}{*}{esc} & $0.0$ & $0.02$ & $2.8 \cdot 10^3$ & $1.86$ & $0.88$ & $63$ & \multirow{3}{*}{$0.343$} & \multirow{3}{*}{$2.27 \cdot 10^3$} \\
 & & $0.349$ & $0.14$ & $400$ & $5.89$ & $0.98$ & $57$ & & \\
 & & $\pi/4$ & $0.272$ & $205$ & $92.74$ & $0.99$ & $56$ & & \\ \hline
\multirow{6}{*}{$\mathbb{S}9$} & \multirow{3}{*}{in} & $0.0$ & $0.03$ & $180$ & $0.27$ & $0.26$ & $21$ & \multirow{3}{*}{$0.335$} & \multirow{3}{*}{$24$}\\
 & & $0.349$ & $0.018$ & $305$ & $0.38$ & $0.36$ & $15$ & & \\
 & & $\pi/4$ & $0.051$ & $108$ & $0.88$ & $0.66$ & $8$ & & \\
 \arrayrulecolor{lightgray} \cmidrule{2-10} \arrayrulecolor{black}
& \multirow{3}{*}{esc} & $0.0$ & $0.1$ & $560$ & $0.54$ & $0.47$ & $119$ & \multirow{3}{*}{$0.348$} & \multirow{3}{*}{$2.24 \cdot 10^3$}\\
 & & $0.349$ & $0.018$ & $3.1 \cdot 10^3$ & $0.38$ & $0.36$ & $155$ & & \\
 & & $\pi/4$ & $0.049$ & $1.1 \cdot 10^3$ & $0.63$ & $0.53$ & $105$ & & \\
\end{tabular}
\caption{\label{table_estimates} Summary of mean transport properties estimates for energetic particles obtained 
with different methods. 
For each simulation the infall (``in'') and escape (``esc'') properties are determined. The infall transport covers a 
distance $ \Delta r_{\rm in} = 5.5 \; r_g$ (from the injection position at $r=8\; r_g$ to the artificial sink $r = 
2.5 \; r_g$). 
The escape properties are determined over a distance $\Delta r_{\rm esc} = 56 \; r_g$ (from the injection position at 
$r=8\; r_g$ to the outer boundary of the region of interest $r = 64 \; r_g$). 
Inflow and outflow could not be distinctly separated at all locations. For regions exhibiting a single behaviour, the 
sole available value was adopted.
$\mathsf{v}_r$ is the radial component of the fluid 3-velocity, from which the ``advection'' estimate 
$\tau_{\rm adv, in} =\Delta r_{\rm in}/v_r$ is derived. 
For the outward (``esc'') directed properties, this velocity reflects the average wind velocity in the funnel-sheath 
region, being used to e.g. evaluate $\tau_{\rm adv, esc} = \Delta r_{\rm esc}/v_r$.
The estimate for the free-streaming time-scales $\tau_{\rm fs} = \Delta r/w_r$ are obtained from the initial ratio 
of radial to azimuthal velocity of the fluid $\mathsf{v}_{r}/\mathsf{v}_{\varphi}$, which determines the radial component 
of the particle $\mathsf{w}_r$. The diffusive parameters $\nu_{\rm in}$, $\nu_{\rm out}$ are obtained as the averages 
of the values in Tab. \ref{table_trapping}, using values for $R \leq 20 \; r_g$ for $\nu_{\rm in}$ and values for $R \geq 
20 \; r_g$ for $\nu_{\rm out}$, and used to evaluate $\tau_\nu = (\Delta r)^2/4\nu$. The angle $\theta = 0.349$  
corresponds to $\theta = 20^\circ$.
}\end{ruledtabular}
\end{table}

\subsubsection{Particle Acceleration} \label{sec_acceleration}
\begin{figure}
\centering
\begin{subfigure}{.32\textwidth}
    \includegraphics[width=\textwidth]{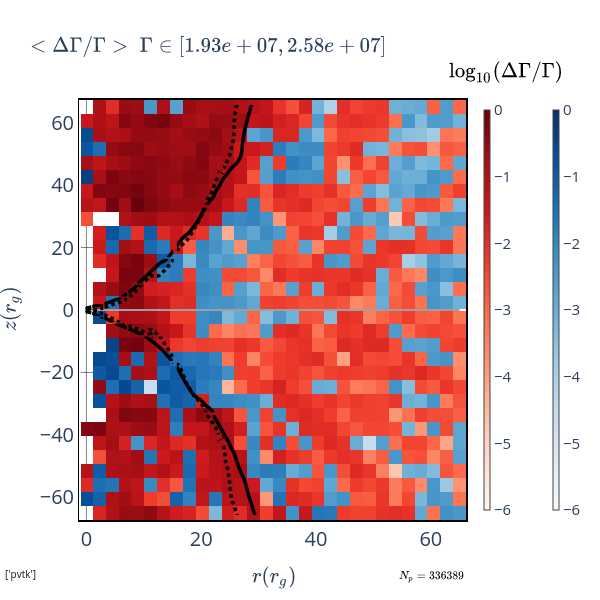}
    \caption{\label{fig_m99_accel_65} $\mathbb{M}9$, analysis done in $R \in [0,64] \; r_g$, $ z \in [-64, 64] \; r_g$.}
\end{subfigure}\,
\begin{subfigure}{.32\textwidth}
    \includegraphics[width=\textwidth]{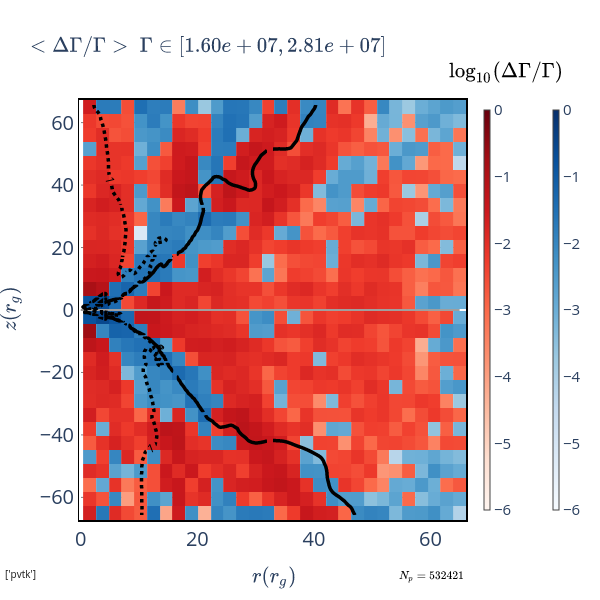}
    \caption{\label{fig_m00_accel_65} $\mathbb{M}0$, analysis done in $R \in [0,64] \; r_g$, $ z \in [-64, 64] \; r_g$ .}
\end{subfigure}\,
\begin{subfigure}{.32\textwidth}
    \includegraphics[width=\textwidth]{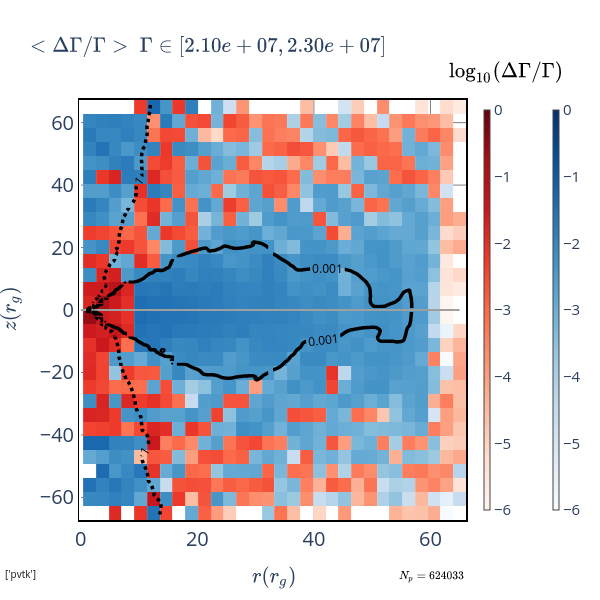}
    \caption{\label{fig_s99_accel_65} $\mathbb{S}9$, analysis done in $R \in [0,64] \; r_g$, $ z \in [-64, 64] \; r_g$ .}
\end{subfigure} 
\caption{\label{fig_accel} Azimuthal average of the average energy change for particles entering and leaving a specific
cell. 
Since the magnitude is shown as the $\log_{10}$ of its value, the red (blue) colorbars indicate positive (negative) values 
of the change in $\Gamma$. The dotted and solid lines indicates the contour levels $\sigma = 1$ and $\rho = 10^{-3}$ 
respectively, highlighting the role played by regions with strong gradients and turbulence in the acceleration of 
particles.
}
\end{figure}

\begin{figure}
\centering
\begin{subfigure}{\DefFigWidth}
    \includegraphics[width=\textwidth, trim = 0 0 0 10em, clip]{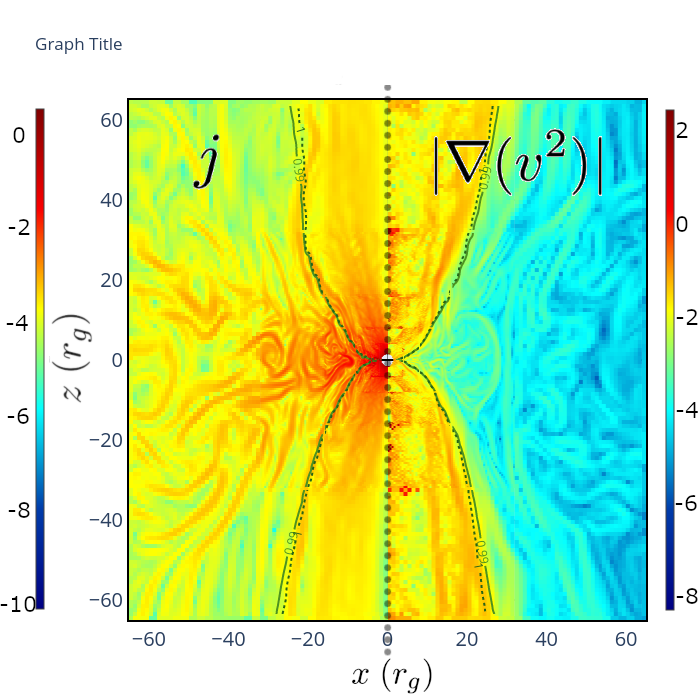}
    \caption{\label{fig_maps_m99}$\mathbb{M}9$}
\end{subfigure}
\begin{subfigure}{\DefFigWidth}
    \includegraphics[width=\textwidth, trim = 0 0 0 10em, clip]{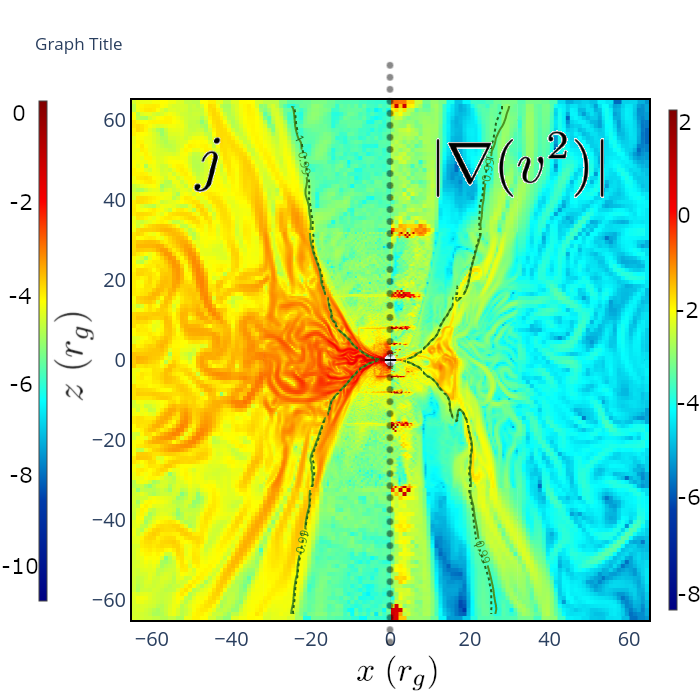}
    \caption{\label{fig_maps_m00}$\mathbb{M}0$}
\end{subfigure}
    \caption{\label{fig_maps_combined} Map in the $x-z$ plane of the base $10$ logarithm of absolute value 
    of the gradient of the fluid velocity magnitude (right half) and of the current (left half). 
    Overlaid with a solid line is the contour for the Bernoulli number $Be = -u_0 \, h/\rho - 1 \simeq 1$, 
    and with a dotted line for the magnetization $\sigma =1$. 
    These contours are nearly coincident in most cases, motivating our use of $\sigma$ alone to identify
    the funnel region in the remainder of the paper.
    Localized maxima in the funnel ($\sigma \geq 1$) region are due to the density floor used to treat 
    the vacuum regions. The left colorbar refers to the current, the right one to the velocity gradient.}
\end{figure}

\noindent \textbf{Acceleration behavior in different accretion regimes}\\
In discussing particle acceleration, we emphasize that the particles considered here are purely passive tracers and thus 
exert no feedback on the disc. While a more complete treatment would be desirable, the results already provide valuable 
insight into the underlying acceleration processes.\\ 
In analyzing the acceleration, we rely on $\Gamma = \alpha u^0$, the Lorentz factor measured by the local FIDO/normal 
observer. This quantity is easily accessible from the code and in principle well suited to describe the local radiation 
physics. However, it has the drawback of depending on the metric explicitly, thus the values reported closest to the 
compact object require care when being interpreted. 
The analysis follows each particle as it travels through the disk and records the variations of its Lorentz factor 
$\Gamma$ as it enters and leaves the cells of the post-processing grid (though motion in the azimuthal direction 
is neglected, i.e., only crossing cells in the $R$ and $z$ directions is considered), thus obtaining a picture of which 
regions of the disk result in larger variations of the particle energy, and in which ``direction'' (acceleration or 
deceleration).\\
As might have been expected, the presence of more strongly magnetized regions, leads to a noticeable localization and 
intensification of the interaction of the particles with the fluid (through the electromagnetic fields generated by its 
dynamics), clearly shown in Fig. \ref{fig_accel}, in the region around the $\rho = 0.001$ contour, encompassing the 
region of the funnel, across all simulations. 
We note here that these regions are where the transition from the disk to the low-density zone at the center of the 
funnel occurs. In this area, the numerical density and pressure floors ($\rho_{fl} = 10^{-10}$, $p_{fl} = 3.3 \cdot 
10^{-12}$) required for the GRMHD computations — which become particularly relevant in the innermost part of the funnel 
\citep{mignone2007equation, ripperda2019general, porth2019event, mizuno2024general} — may influence the structure of 
the flow. In particular, they could affect the magnitude and spatial extent of the gradients in the fluid quantities.
Consequently, while the qualitative trends seem robust, their precise quantitative characterization may depend, 
at least in part, on the adopted floor prescription.
The region where the floors impact the fluid quantities is visible in Fig. \ref{fig_maps_combined} as the noisy 
central part of the velocity gradient.\\
Going into detail for each of the simulations run, $\mathbb{M}9$ shows the largest relative acceleration (or 
deceleration), both closer and further from the BH, with increasing intensity along the funnel region. 
The disk can still affect particles' energies, though its importance is secondary to 
the funnel region. Particles are unlikely to stay in the innermost regions of the funnel as they propagate 
outwards, and instead move along the region of transition between the funnel and the disk, where most of the 
acceleration takes place. 
It is also worth noting that, although the magnetization attains its maximum in the funnel core, the magnetic 
field strength peaks close to the denser funnel sheath, which may help to explain the localization of the 
acceleration.\\  
$\mathbb{M}0$ shows similar behavior, though the innermost region of the funnel seems to be less important than 
in $\mathbb{M}9$, possibly due to the overall lower values of magnetic field reached when the BH has lower spin. 
However, the relative acceleration achieved by particles is lower than in the case with larger spin, which suggests 
a relevant role played by the magnitude of the magnetic field (and its fluctuations) in the acceleration mechanism. 
Crucially, however, the behavior close to the BH, is very different between the two MAD cases: 
$\mathbb{M}9$ shows interaction between particles and disk inside the funnel region close to the BH, 
while $\mathbb{M}0$ exclusively shows interaction outside the $\sigma \geq 1$ region. 
Associating this behavior with the dynamics of the accretion disk across the two simulations may suggest that 
the stronger magnetic field and outflows in the $\mathbb{M}9$ case close to the BH play a role in the 
localization of the interaction. Alternatively, it is possible that there is a common acceleration mechanism at 
play in the region where the funnel (the jet in $\mathbb{M}9$) transitions to the disk (the sheath), 
and an additional one inside the jet of $\mathbb{M}9$.\\
Indeed, Fig. \ref{fig_s99_accel_65} shows that even in $\mathbb{S}9$ the funnel region is the seat of intense 
interactions between particles and fluid, though $\mathbb{S}9$ is the least efficient accelerator of the cases 
analyzed.\\
These acceleration regions are most effective for particles at the lower end of the energy spectrum, whereas 
particles with higher energies experience almost exclusively decreases in energy.\\ 
We remind that changes in $\Gamma$ in the region closest to the BH can be attributed to the role of the 
metric, such that the only effect visible in a simulation run (not shown here) for a SANE case with $0$ spin 
is the change in $\Gamma$ due to the particles leaving 
the gravitational potential (with an almost exact dependence $\sim 1/r$ of the energy change).\\ 
We also note the ``arch-like'' features in the localization of the acceleration in Fig. \ref{fig_s99_accel_65}, 
that match rather well the regions of strong $\mathsf{v}$ gradient in Fig. \ref{fig_sigma}, and are a strong 
indication that the acceleration mechanism, also in the disk, is strongly correlated with gradients in the 
properties of the fluid. Note that the MAD cases show negligible acceleration features in the denser regions 
of the disk, a difference that might be attributable to the different flow structures in MAD and SANE disks,
as the SANE cases are much closer to an axisymmetric flow (\citep{narayan2003magnetically, narayan2012grmhd}), 
that can thus act ``consistently'' (systematically), rather than stochastically, on the particles.\\
Indeed, in the absence of fluid-particle interactions, $\Gamma$ would simply decrease as particles move 
outward due to receding gravitational time dilation and kinetic deceleration, explaining the large regions 
of average negative $\langle \Delta \Gamma \rangle$, especially at higher energies, where interactions with 
the fluid become less frequent. Only when interactions are frequent enough does 
$\langle \Delta \Gamma \rangle$ turn positive, as is e.g. the case in the funnel sheath regions of all the 
simulations. At the same time, this means that one needs to ensure that a sufficient number of particles are involved in the 
averages, as the negative changes in $\Gamma$ from free-streaming are the default behavior for most particles, especially at high energies.
Alternatively, the differing degrees of magnetization (see also Fig. \ref{fig_sigma}) may be the key factor 
determining the distinct acceleration behaviors, a point that will be further discussed at the end of this 
section. 
It is possible that the low efficiency of $\mathbb{S}9$ in accelerating particles stems from the quick 
transition of particles to the free-streaming regime, which could be mitigated by choosing different injection 
energies, with all the challenges, mentioned in Sec. \ref{sec_setup} that come with resolving those orbits 
reliably across the whole domain.\\
Tentative extrapolation would suggest that particles with even smaller Larmor radii would then be more 
efficiently accelerated by the fluid dynamics in all regimes, though this remains to be verified.
Still, the results presented here show that even when particles have reached significantly larger energies 
than the background fluid, both MAD and SANE disks can continue to accelerate them further. This, in turn, 
suggests that the distribution of energetic particles is not only shaped by local injection and acceleration 
mechanisms, but also by their longer term dynamics and interaction with the disk, as will be explored 
further in Sec. \ref{sec_individual}.\\
The average distribution over the whole domain over all times for $\mathbb{M}9$ is shown in Fig. \ref{fig_avg_dist} 
to form an extremely hard tail, through dynamics that are explored more in detail in Sec. \ref{sec_individual}. 
The tail can be fit with a power-law index close to $\xi \approx 1$ (see Fig. \ref{fig_avg_dist}, which 
is compatible with previous studies \citep{kimura2015llagn, kimura2019acceleration}, although in the present 
study we use a fully GR system and our results have the strong limiting factor that the injection is not 
continuous, and thus does not establish a proper steady-state condition.\\
This type of analysis could be used as a method to validate predictions of spectral shapes from 
theoretical arguments (e.g., \citep{lemoine2025neutrinos}), although to carry out this task more 
accurately a mechanism for continuous injection of particles and feedback of the particles on the 
fluid are crucial components that limit the applicability of the results shown here.\\

\begin{figure}
\centering
\begin{subfigure}{\DefFigWidth}
\includegraphics[width=\textwidth, trim = 0 0 0 0, clip]{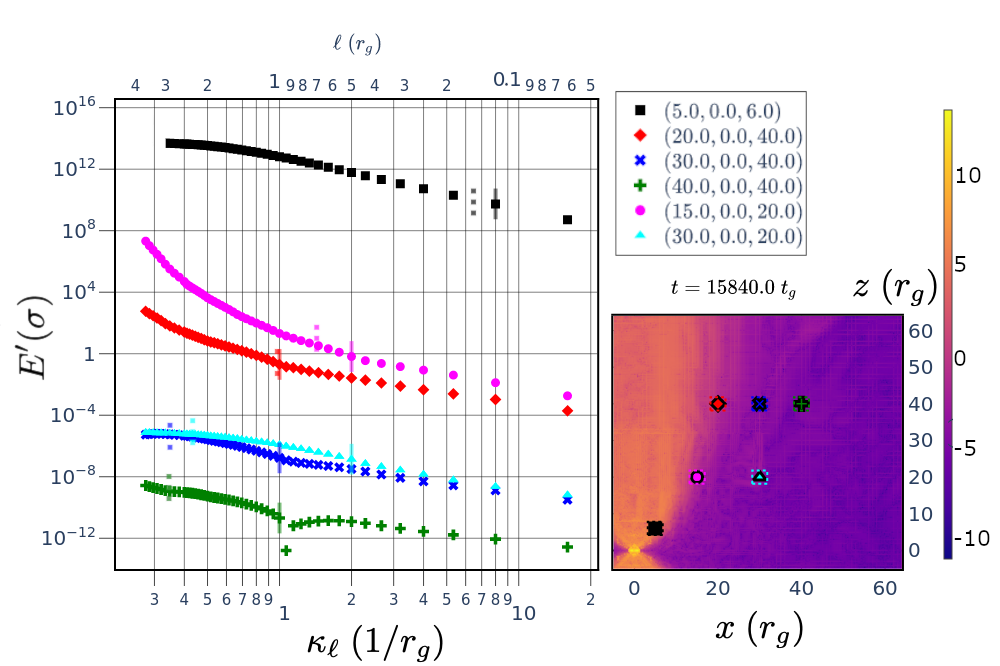}
\caption{\label{fig_spectra_m9}$\mathbb{M}9$}
\end{subfigure}
\begin{subfigure}{\DefFigWidth}
\includegraphics[width=\textwidth, trim = 0 0 0 0, clip]{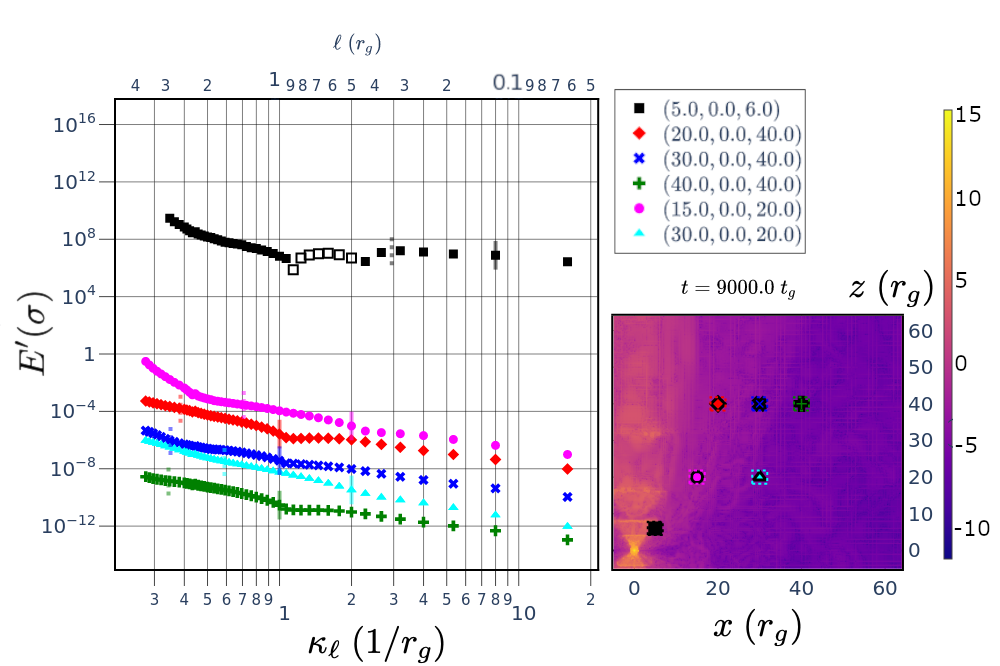}
\caption{$\mathbb{M}0$}
\end{subfigure}
\caption{\label{fig_spectra} High-pass filter spectra of the magnetization at the locations $(x,y,z)$ 
indicated in the heatmaps at the bottom right of each sub-figure, for simulation $\mathbb{M}9$ and $\mathbb{M}0$. 
The heatmaps show the $\ell=1$ high-pass filtered field. The dashed vertical lines show the Larmor radii for 
particles with $\Gamma = 10^8$ at each location, while the solid vertical lines show the computational grid 
size at that position. The squares in the bottom right portion of the figures indicate the maximum extent of the 
coarse-graining around a given point.}
\end{figure}

\noindent \textbf{Understanding the acceleration dynamics}\\
The two key candidates to explain the observed acceleration behavior are intense local electric fields 
from magnetic reconnection \citep{ripperda2019general, zhang2021fast, ripperda2022black, dihingia2025impact} 
and scattering on structures of the magnetic field, whether it be turbulence, curvature or shear 
\citep{lemoine2019generalized, rieger2019,wang2021particle, lemoine2023particle}. 
In the present case, reconnection is due to grid-scale numerical dissipation, since no resistive terms 
are implemented. Distinguishing the two mechanisms can prove difficult, since the currents required to 
drive reconnection can be due to turbulence, or be the result of an enhancement of a local shearing of 
the magnetic field by the flow dynamics, while the consequence of magnetic reconnection can be an 
enhancement of turbulence. Exploring which of the two is at play here will be the goal of the remainder 
of the section.\\
Fig. \ref{fig_maps_combined} show the current density and the velocity gradient distribution in the system. 
Since the areas we are focusing on are far from the compact object, we use the non-relativistic definition 
and compute the current as $\bf{j} = \nabla \times \mathbf{B}$ (i.e., we neglect the displacement current), 
and the gradient as $\nabla_i = \partial_i$. In principle, computing the current this way mainly provides 
a visualization of irregularities in the magnetic field.
As the maps are obtained from data averaged over $\Delta t \geq 100 \; t_g$, they do not capture the 
effects of transient reconnection events. However, one can expect that most reconnection sites will 
coincide with regions of strong average currents, reflecting the fact that magnetic reconnection is 
due to current sheaths (usually already present) becoming excessively elongated due to the dynamics of 
the flow. Such current sheath are then ``converted'' to electric fields capable of accelerating particles 
by the resistivity. 
It is noticeable that the region with the most intense current density is the inner ($R \leq 30 \; r_g$) 
mid-plane, and its localization aligns well with the disk regions highlighted by the analysis in Fig. 
\ref{fig_m00_accel_65} and even more Fig. \ref{fig_s99_accel_65}, where significant acceleration takes 
place in the areas of the disk closest to the BH. \\
However, looking more closely at Fig. \ref{fig_maps_m99}, one notices that the current in the jet region 
(enclosed by the isocontours $\sigma = 1$) tends to weaken away from the BH, while the region of 
acceleration within the jet in Fig. \ref{fig_m99_accel_65} seemed to be rather uniform where present, 
though we stress that the innermost part of the jet is prone to numerical artefacts, and drawing 
conclusions from what is observed in this region should be done cautiously. 
Crucially, if the main acceleration mechanism were currents related to magnetic reconnection, one should 
not expect a threshold in the energy of the particles where these stop being accelerated and are instead 
decelerated. If anything, one could expect a dependence of the acceleration and deceleration on the region 
of the domain, but constant across particle energy, which is not observed here. Finally, if reconnection 
were the main driver of the acceleration, one should expect, on average a change in the particle energy 
of the order $\sim \sigma$, which can be seen not to be the case by comparing the maps in Fig. \ref{fig_accel} 
and Fig. \ref{fig_sigma}.\\ 
The energy dependence of acceleration and deceleration is much more consistent with the alternative explanation 
of scattering: particles with lower energies than that available in the turbulence, or whose Larmor radius 
is one the same scale as the structures in the magnetic field will, on average, gain energy by interaction with 
the field, whereas particles with higher energies will lose energy (even though in these simulations we do not 
actually subtract or deliver this energy from and to the fluid). This latter point is of interest also for 
future work, where the feedback of particles on the disk could be included in the simulations, and high-energy 
particles losing energy to the fluid could modify the dynamics on larger scales.
Finally, from the spectra of the magnetic field fluctuations (see App. \ref{sec_coarse_graining}) 
one can see how further from the BH the fluctuation level quickly exceeds $\sqrt{\delta B^2/ \langle B\rangle^2} > 1.0$,
indicating much stronger fluctuations at the same locations where the acceleration intensifies. 
Considering these points, and the ones that will be further provided in Sec. \ref{sec_individual}, 
the possibility that magnetic reconnection is the main 
source of acceleration seems rather unlikely. The currents shown in Fig. \ref{fig_maps_combined} are then a useful way 
of visualizing and quantifying these structures in the magnetic field, rather than a means to look into reconnection.\\
Interaction of the particles with the dynamical structures of the magnetic field is the main alternative explanation 
for the acceleration behavior. In particular, the region of the funnel-sheath is, 
as already mentioned, a region of strong and chaotic velocity gradients (see Fig. \ref{fig_maps_combined}), 
and can thus drive instabilities like the KH mode, that 
in turn result in turbulent structures that can interact with the particles, leading to rich dynamics, and, through 
scattering, to stochastic acceleration \citep{lemoine2019generalized, jafari2019topology, kimura2019acceleration, 
bentkamp2019persistent,wang2023KHI}. 
Another strong indication that turbulence is the main acceleration mechanism is the similarity in the interaction 
patterns in Fig. \ref{fig_s99_accel_65} (they look like spherical shells localized around the mid-plane, that become 
more or less visible depending on the binning on the particles) and the structure of the inflow (radial $\mathsf{v}_r$) 
velocity shown in Fig. \ref{fig_sigma}). The acceleration localizes strongly in regions where the gradient of the 
inflow velocity is the strongest. Thus the localization of the acceleration regions in Fig. \ref{fig_accel}, together 
with the analysis presented in Fig. \ref{fig_maps_combined}, provides the first indication that turbulence is the 
primary acceleration mechanism.\\
In order to explore this further, we perform coarse-graining \citep{jafari2019topology,zhao2025calculating} 
on the MHD fluid, even though the dynamic range of the resulting spectrum is limited in range. 
The need to use coarse-graining arises from the strongly anisotropic and inhomogeneous nature of the system we are 
analyzing, that also lacks clear periodic directions, where Fourier modes work best. 
The details of the coarse-graining can be found in App. \ref{sec_coarse_graining}, here we mention the definition 
of the fluctuation amplitude for the magnetization $E'(\sigma)$ and the ``wave-number'' 
at scale $\ell$, $\kappa_{\ell} \equiv \frac{L}{\ell}$.\\
\begin{equation}
E'(\sigma) \equiv \frac{d}{d k_\ell} \left\langle \frac{1}{2}\vert \sigma'_\ell (x) \vert^2 \right\rangle_X = - \frac{\ell^2}{L} \frac{d}{d\ell} \left\langle \frac{1}{2}\vert \sigma'_\ell (x) \vert^2 \right\rangle_X
\end{equation}
Here, $L$ is a typical scale of the system, and the ``high-pass-filtered'' magnetization $\sigma'$ 
is obtained from the coarse-grained $\bar{\sigma}$ through $\sigma' = \sigma - \bar{\sigma}$.\\
The spectra shown here are obtained with isotropic convolution, without distinguishing particular directions.\\
Coarse-graining allows to extract the spectra shown in Fig. \ref{fig_spectra}. 
Comparing these spectra to Fig. \ref{fig_maps_combined}, Fig. \ref{fig_m99_accel_65} and \ref{fig_m00_accel_65}, 
one can notice how they can easily explain the localization of the acceleration: 
the regions where the turbulence is stronger in Fig. \ref{fig_spectra_m9} (i.e., at locations $(15, 0, 20)$ and $(20, 0, 40)$, 
where in this notation the numbers refer to the cartesian coordinates $(x,y,z)$ of the point and no longer to the cylindrical ones) 
match rather well the most intense acceleration regions along the jet-sheath. 
Furthermore, regions with similar turbulent spectra across 
the two MAD cases (i.e., spectra at $(30, 0, 40)$ and $(30, 0, 20)$) are also the regions that showed similar acceleration 
behavior in Fig. \ref{fig_accel}, whereas the current was varying significantly (Fig. \ref{fig_maps_combined}) in regions where the 
acceleration behavior was similar. Note that the funnel region is known to be the seat of intense turbulent heating across SANE 
and MAD cases \citep{dihingia2023temperature}, providing further strength to our interpretation.\\
The different behavior in the funnel region closest to the BH can be understood by noting that in the funnel region 
the magnetic field in $\mathbb{M}0$ is much weaker than in the disk, which is not the case in $\mathbb{M}9$. 
As a results, the Larmor radii in the central region of $\mathbb{M}0$ are excessively large, preventing particles 
from interacting with the turbulent structures, a further indication
that turbulence is the driver of the acceleration. At the same time, this suggests that the choice of $\sigma$ as the quantity for 
coarse-graining might not have been optimal, or that the normalization of the spectra could warrant more careful considerations. 
Nevertheless, we believe that the analysis presented offers a satisfactory explanation of the dynamics.\\
The statistical analysis presented here already provides a strong and general understanding of the underlying processes, 
suggesting that particle dynamics are driven more by large-scale turbulent field structures than by magnetic reconnection. 
Further evidence for this picture will be discussed in Sec.~\ref{sec_individual}.\\
If turbulent scattering is considered as the dominant mechanism for particle energization, it is possible to correlate 
the behavior of the energy change to the underlying process that causes such changes. Given the location of the most 
intense interaction in Fig. \ref{fig_accel}, two candidates emerge: (classical) second-order Fermi acceleration and
flow-shear acceleration. The distinction can be probed via the expected relative energy change $\Delta E/E$. For 
second-order Fermi acceleration $\Delta E/E \sim \Gamma_w^2 v_A^2 \sim \sigma$, where $v_A = \alpha B/\sqrt{h}$ is the 
Alfv\'en velocity, $h = \rho + \chi \epsilon + (\alpha B)^2/2 $ is the enthalpy for a magnetized gas (computed using 
the adiabatic index $\chi = 4/3$ for a perfect relativistic fluid) and $\epsilon$ is the internal energy. 
For flow-shear acceleration \citep{rieger2004, rieger2019,wang2021particle} $\Delta E/E \sim \Gamma_w^4 (\Delta w)^2$, 
where $\Gamma_w$ is the gamma factor of the flow. One can immediately notice how the quick increase in $\sigma$ visible 
in Fig. \ref{fig_sigma}, which can even span several orders of magnitude going from the sheath to the center of the 
funnel, does not correspond to a significantly enhancement $\Delta \Gamma/ \Gamma$ in Fig. \ref{fig_accel} in the 
central part of the funnel region compared to the sheath. If anything, intense acceleration, indicated by the darker 
red colors in Figs. \ref{fig_m99_accel_65} and \ref{fig_s99_accel_65} occur along the $\sigma = 1$ contour, that 
Fig.~\ref{fig_maps_combined} identifies as the region of strong gradients in the fluid velocity. 
One can also notice how in Fig. \ref{fig_maps_m00} the most intense gradients of the velocity (with the exception 
of regions affected by numerical floors in the center of the funnel), are localized at the mid-plane around
$R \approx 15 \; r_g$, which matches very well the region of most intense acceleration in Fig. \ref{fig_m00_accel_65}. 
The picture of flow-shear-driven acceleration also removes the need for different acceleration processes acting 
inside the jet of $\mathbb{M}9$ and in the sheath region of both $\mathbb{M}9$ and $\mathbb{M}0$ discussed  
earlier in this section, as the behavior of the velocity gradient can explain the difference between these 
cases. 
With turbulence as driving mechanism, a cohesive picture emerges of how different accretion regimes 
affect particle acceleration. In $\mathbb{M}9$, the powerful jet driven by the Blandford-Znajek (BZ) 
\citep{blandford1977electromagnetic} provides the strongest turbulence in the sheath region, where 
the velocity field accelerates to relativistic speeds all the way to the  center of the jet itself. In 
$\mathbb{M}0$, the BZ process is inactive due to the BH's zero spin, and the weaker winds from the 
Blandford-Payne (BP) mechanism \citep{blandford1982hydromagnetic} act as the source of the velocity 
gradient. This gradient is not only weaker but also does not extend to the BH's pole, explaining the reduced 
acceleration in the innermost funnel region. Consistently, $\mathbb{S}9$ shows weaker acceleration: 
although the BH has high spin, the available magnetic flux is insufficient to drive a strong jet like that
in $\mathbb{M}9$, and lower accretion rates fail to energize BP winds to the same level as in $\mathbb{M}0$, 
though they still generate mild turbulence in the sheath. 
The weaker confinement of high-energy particles in $\mathbb{S}9$ further limits acceleration.
These insights may also inform our understanding of other accretion regimes exhibiting shear-driven 
dynamics, such as the hydrodynamic ADIOS model \citep[e.g.,][]{blandford1999fate, begelman2012radiatively, 
abramowicz2013foundations}, where mid-plane inflow transitions to outflows at higher latitudes. However, 
since the velocity shear layers in ADIOS are expected to be slower and weaker than in the magnetized cases 
considered here, the underlying acceleration mechanisms might be modified, warranting dedicated investigation.\\
Finally, as already remarked, using the radial velocity as a proxy for the total velocity (Tab. \ref{table_estimates} 
justifies this as an order-of-magnitude estimate, at least in the innermost parts of the system), 
Fig. \ref{fig_sigma} shows that $\mathbb{S}9$ reveals rather uniform and stable in time 
gradients of the velocity (the arch-like structure that were already highlighted in Sec. \ref{sec_escape}) 
through the whole disk. This correlates very well with the behavior of the acceleration shown 
in Fig. \ref{fig_s99_accel_65}, as previously noted. Together, these results strongly suggest that particle dynamics 
are primarily governed by scattering against turbulence driven by flow shear.

\subsubsection{Particle Trapping} \label{sec_trapping}
Particles occasionally become temporarily trapped in localized regions, where repeated scattering reverses or 
randomizes their motion over short timescales, effectively confining them for extended periods. This behavior 
is relevant for assessing how specific system phenomena influence particle dynamics.
For example, if a phenomenon lasts $\sim 10 \; t_g$ but particles remain in any given region for only 
$\tau_{\rm trap}\sim 1$, its impact is limited. Conversely, if the phenomenon occurs on a timescale 
$\sim 0.1 \; t_g$, its effect can be significantly amplified for particles trapped longer in the region.\\
In order to assess this behavior, over the whole trajectory of a particle, its position is mapped to the 
post-processing grid, and the time $\tau_{\rm trap}$ each particle spends in a given region of size 
$\Delta r = 2.1 \; r_g$, $\Delta z = 4.3 \; r_g$ is used to compute an average, that allows to observe regions 
of the disk that are more effective at trapping particles. 
At the same time, this can be combined with the previous (Sec.~\ref{sec_acceleration}) diagnostic to get 
the quantity $\tau^{-1}_{acc} = \frac{\Delta\Gamma}{\Gamma}\frac{1}{\tau_{\rm trap}}$, that quantifies the 
efficiency of a given region in accelerating (or decelerating) particles. 
Not considering azimuthal motion is particularly significant for this diagnostic, as the fluid, far from the 
BH, moves primarily azimuthally, which means that if advection alone were responsible for particle motion, 
this could result in very large $\tau_{\rm trap}$. 
This information is also needed to evaluate the nominal diffusion coefficient in the poloidal 
plane as a function of position (not shown, maximum values are reported in Tab. \ref{table_trapping}), 
as defined in Eq. \ref{eq_diff_coeff}: 
$ \nu = ((\Delta R)^2 + (\Delta z)^2) / (2 \; \eta \; \tau_{\rm trap })$.\\ 
The physical explanation for the trapping will be examined in detail in Sec. \ref{sec_individual}, but a trend emerges with
the highest average trapping times occurring in the innermost regions ($R \leq 20 \; r_g$), which points 
again to scattering by turbulence. 
Importantly, if the main mechanism impacting the dynamics were magnetic reconnection, there should be an 
opposite relation between acceleration and trapping, as the reconfiguration of the magnetic field nulls 
occurring at the reconnection sites are at the same time accelerating and ``freeing'' particles from their 
gyration constraint. The localization of strong trapping in high-field regions suggests such zones will 
play an even greater role in simulations with a corona-like structure, where regions of strong 
magnetization can form also at lower koloidal angles than the funnel region \citep{zhang2025radiation}.\\
The absolute maxima of the trapping time reported in Tab. \ref{table_trapping} all occurred in the outer 
parts ($R \approx 40 \; r_g$) of the region of interest. 
Analysis in the larger domain ($R \leq 500 \; r_g$ and $|z| \leq 500 \; r_g$ shows the dynamics reaching 
$<\tau_{\rm trap}> \approx 15 \; t_g$ over narrow portions of the domain (roughly following $R \sim |z|$) for 
all simulations, and up to $\tau^{max}_{\rm trap} \geq 300 \; t_g$ in the MAD cases, though the latter value 
can be directly linked to the initial conditions of the simulation, as it occurs at $R \approx 150 \; r_g$ 
which is (in the mid-plane) where the Fishbone-Moncrief torus ``ends'', meaning that the magnetic field drops 
below $\vert B \vert \leq 10^{-10}$. Just before decreasing below this threshold, it is 
almost purely azimuthal, and since the present analysis disregards azimuthal motion for the trapping time, 
confinement times are ``artificially'' large.\\
The values of the acceleration rate indicate that the typical particle acceleration process takes place on 
time scales much smaller than those on which the MHD quantities evolve (see Fig. \ref{fig_history}). This 
implies that the interaction between the particles and the dynamics of the fluid (e.g., the periodic ejections 
in $\mathbb{M}9$) are going to be limited. 
The difference between simultaneously evolving the fluid and particles and treating the fluid as a static 
snapshot is thus expected to be modest.

\begin{table}
\begin{ruledtabular}
\begin{tabular}{|cccccccc|}
\textbf{Simulation} & $\Gamma $ & $<\tau_{trap}>\vert^{max}_{R \, < \, 20\, r_g} $ & $\nu\vert^{min}_{R \, < \, 20\, r_g} $ & $<\tau_{trap}> \vert^{max}_{R \, > \, 20\, r_g}$ & $\nu \vert^{min}_{R \, > \, 20\, r_g}$ & $\tau^{max}_{trap}$ & $<\tau^{-1}_{acc}> \vert^{max}_{R \, < \, 20\, r_g}$ \\ \hline
\multirow{2}{*}{$\mathbb{M}9$} & $[0.1, 1.5] \times 10^8$ & $13 \; t_g$ & $0.88 \; c\cdot r_g$ & $9 \; t_g$ & $0.32 \; c\cdot r_g$ & $70 \; t_g$ & $0.45 \; t^{-1}_g$ \\
& $[2.0, 3.0] \times 10^8$ & $9.9 \; t_g$ &  $1.16 \; c\cdot r_g$ & $9 \; t_g$ &  $0.32 \; c\cdot r_g$ & $45 \; t_g$ & $0.035 \; t^{-1}_g$ \\ \hline
\multirow{2}{*}{$\mathbb{M}0$} & $[0.1, 1.5] \times 10^8$ & $7.2 \; t_g$ & $1.59 \; c\cdot r_g$ & $7.8 \; t_g$ & $0.37 \; c\cdot r_g$ & $52 \; t_g$ & $0.08 \; t^{-1}_g$ \\
& $[2.0, 3.0] \times 10^8$ & $8.4 \; t_g$ & $1.36 \; c\cdot r_g$ & $9 \; t_g$ & $0.32 \; c\cdot r_g$ & $36 \; t_g$ & $0.025 \; t^{-1}_g$ \\ \hline
\multirow{2}{*}{$\mathbb{S}9$} & $[0.2, 1.5] \times 10^7$ & $9.2 \; t_g$ & $1.24 \; c\cdot r_g$ & $8.2 \; t_g$ & $0.35 \; c\cdot r_g$ & $31 \; t_g$ & $0.08 \; t^{-1}_g$ \\
& $[4,0, 5.0] \times 10^7$ & $8 \; t_g$ & $1.43 \; c\cdot r_g$ & $8.2 \; t_g$ & $0.35 \; c\cdot r_g$ & $12 \; t_g$ & $0.02 \; t^{-1}_g$ 
\end{tabular}
\caption{\label{table_trapping} Summary of different properties for particles in different energy regimes 
related to the analysis within the region $R \leq 64 \; r_g$ and $|z| \leq 64 \; r_g$. $\tau_{\rm trap}$ 
refers to the time a particle stays in one of the cells defined for the post-processing grid 
(cells have size $\Delta r = 2.1 \; r_g$, $\Delta \varphi = \pi/10$, $\Delta z = 4.3 \; r_g$), 
without considering azimuthal motion, i.e., considering only motion that might lead to particle loss.
The acceleration rate is defined as $\tau^{-1}_{acc} = \frac{\Delta\Gamma}{\Gamma}\frac{1}{\tau_{trap}}$. 
The diffusion coefficient is defined as 
$\nu = \frac{(\Delta R)^2 + (\Delta z)^2}{2 \; \eta \; \tau_{\rm trap }}$. }
\end{ruledtabular}
\end{table}

\subsection{Individual Particle Dynamics} \label{sec_individual}
\begin{figure}
\centering
\begin{subfigure}{.45\textwidth}
    \includegraphics[width=\textwidth]{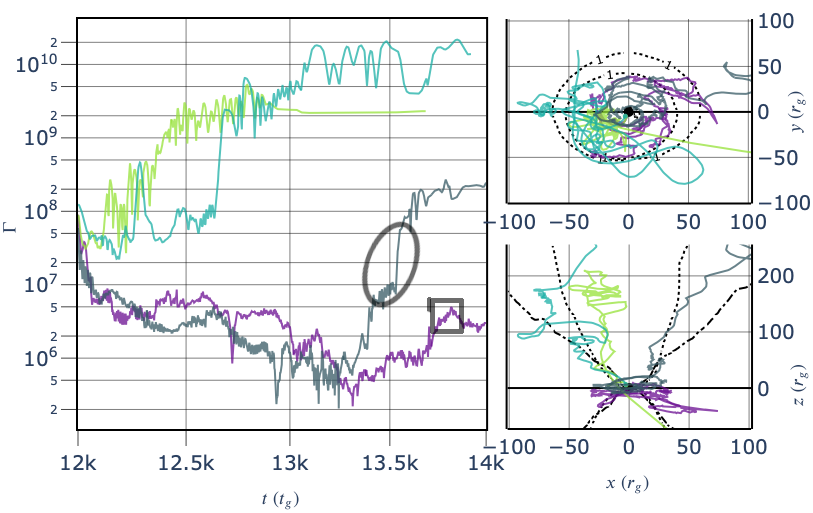}
    \caption{\label{fig_inhst_det} Trajectories (right panels) and evolution over time (left panel) of 
    $\Gamma$ for selected particles. 
    The right panels also shows the contour $\rho = 10^{-3}$ as dash-dotted black line and $\sigma = 1$ as 
    dotted black lines. In the $(x,y)$ plane the $\sigma=1$ contours are shown for $z=75, \, 150, \, 300 \; r_g$. 
    The black square and ellipse represent the particles and time-frames that are expanded upon with the 3D snapshots in Figs.     \ref{fig_turbulence} and \ref{fig_inversion} respectively.}
\end{subfigure}\;
\begin{subfigure}{.45\textwidth}
    \includegraphics[width=\textwidth, trim=0 0 7.5em 11em, clip]{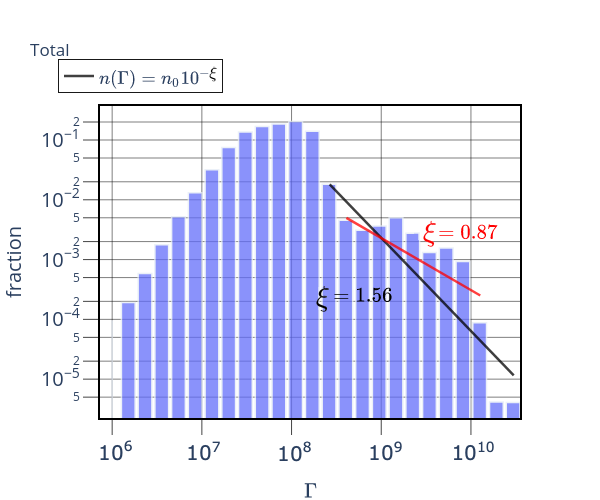}
    \caption{\label{fig_avg_dist} Average particle distribution in the region of interested for our analysis
    ($R \in [0, 64] \; r_g$, $z \in[-64, 64] \; r_g$). 
    The energy axis has been extended with respect to the previous analysis and made with logarithmic spacing. 
    Two different power-law fits for the high-energy tail are shown. }
\end{subfigure}
\caption{Figure highlighting the behavior of individual particles and how they can impact the average properties 
    of the system in $\mathbb{M}9$. The average particle distribution shows a very extended tail into 
    high-energies, due to particles being accelerated along the jet and then reflected back into the region of
    interest.}
\end{figure}

To better understand the processes experienced by the particles, we focus in this section on a subset of 
particles that exemplify the larger sample, and study their behavior in detail. Although more particles 
were analysed, those presented here give an overall picture of the most interesting dynamics occurring in 
the system. These particles are taken from $\mathbb{M}9$.\\
Fig. \ref{fig_inhst_det} shows, in the left panel, the evolution over time of the $\Gamma$ factors of selected 
particles. The right panels show the projection of the trajectories in the $(x,z)$ (bottom panel) and $(x,y)$ 
(top panel) planes.\\ 
Inspection of the trajectories in Fig. \ref{fig_inhst_det} shows that particles undergoing significant energy 
changes are those that experience multiple scattering events, and are thus remaining in the domain for rather 
long times, or those that enter the funnel region. This further supports the interpretation that stochastic 
acceleration is the main mechanism for particle energization, while magnetic reconnection (or, more generally, 
acceleration by resistive electric fields) does not seem to play a significant role, at least outside the 
innermost region ($R \leq 30 \; r_g$) since the currents in the region of interest are not particularly 
intense or localized and their amplitudes decrease as the distance from the BH increases (Fig. \ref{fig_maps_combined}), 
whereas the average energy change can occur with almost constant magnitude over much broader regions. 
At the same time, for higher-energy particles, the largest energy changes take place in between the region 
of the disk where currents are localized and the $\sigma \geq 1$ region starts (and, as already discussed 
in Sec. \ref{sec_acceleration}, stronger velocity gradients are present), thus it is possible that for these 
particles electric fields due to grid-scale reconnection may still play a role.\\
Notably, since, for the most part, particles at such high energies are free-streaming and only occasionally 
interacting with the fluid, changes in the normal-frame Lorentz factor $\Gamma$ reflect changes in the metric,
that have, roughly, a $\sim 1/r$ dependence, same as the magnitude of the magnetic field, ensuring stable resolution 
of Larmor orbits even for apparently large changes in $\Gamma$.\\

\begin{figure}
\centering
\begin{subfigure}{.31\textwidth}
    \includegraphics[width=\textwidth]{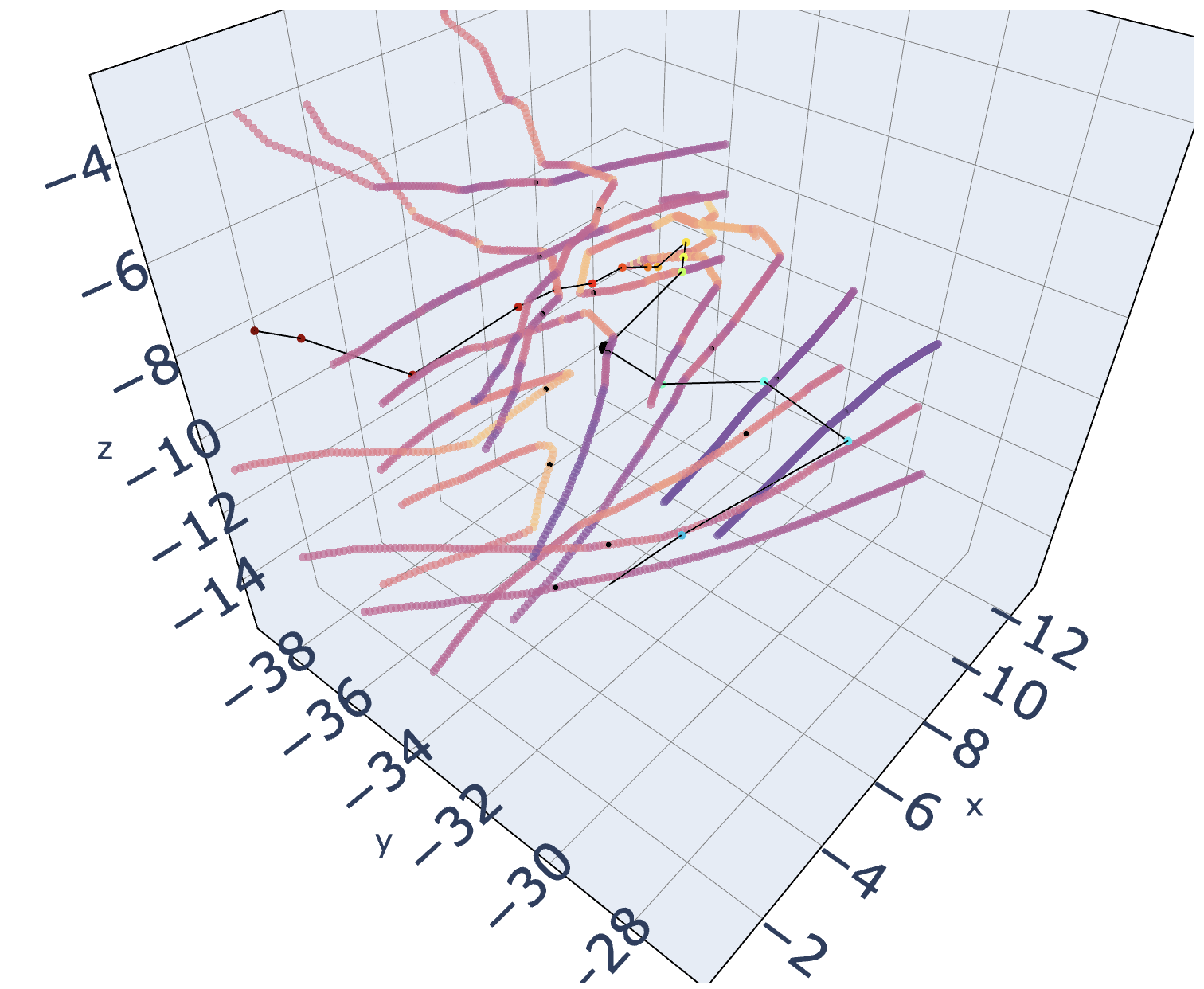}
    \caption{\label{fig_t1} $t = 13755 \; t_g$ }
\end{subfigure}\;
\begin{subfigure}{.31\textwidth}
    \includegraphics[width=\textwidth]{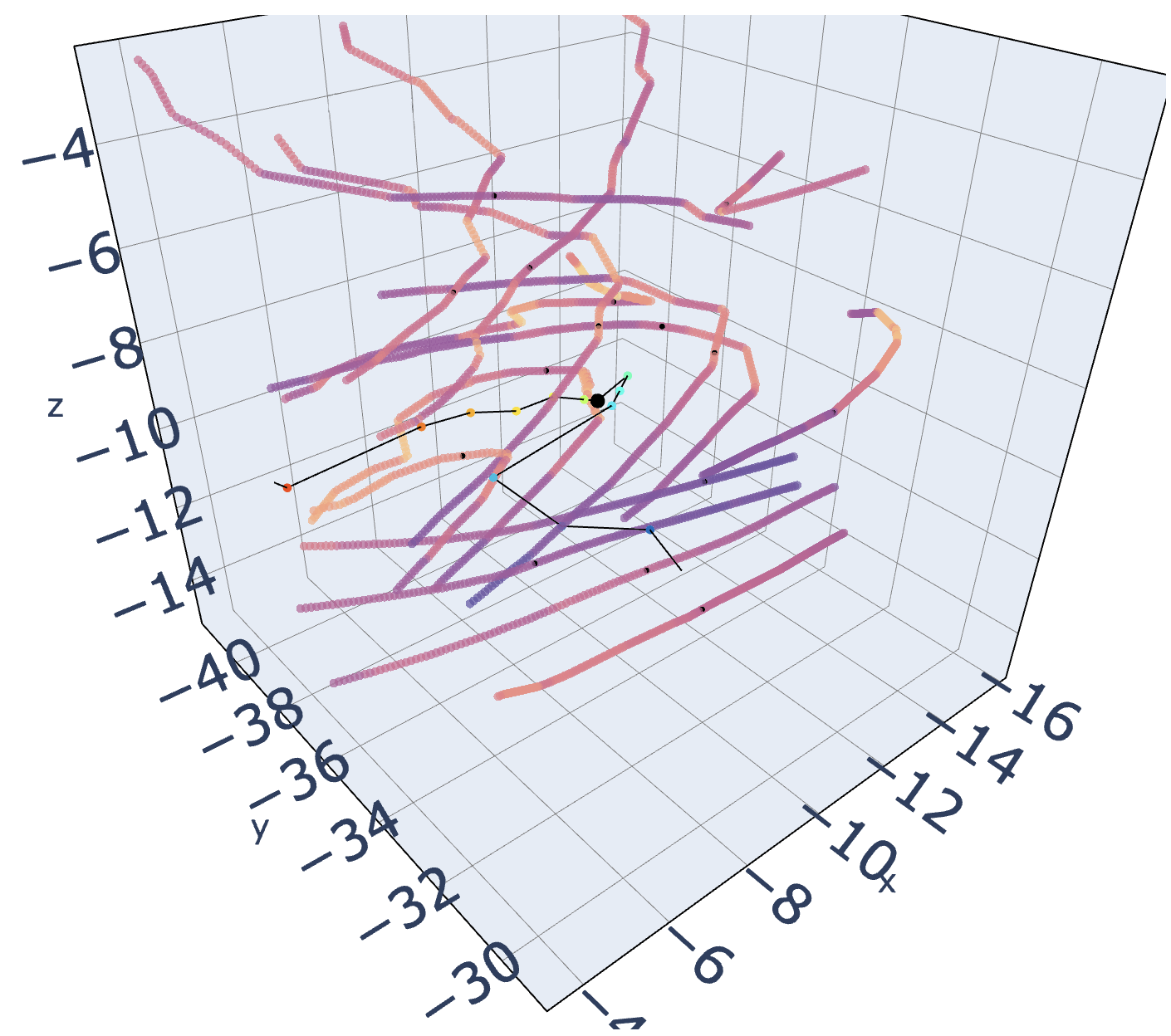}
    \caption{\label{fig_t2} $t = 13773 \; t_g$ }
\end{subfigure}\;
\begin{subfigure}{.31\textwidth}
    \includegraphics[width=\textwidth]{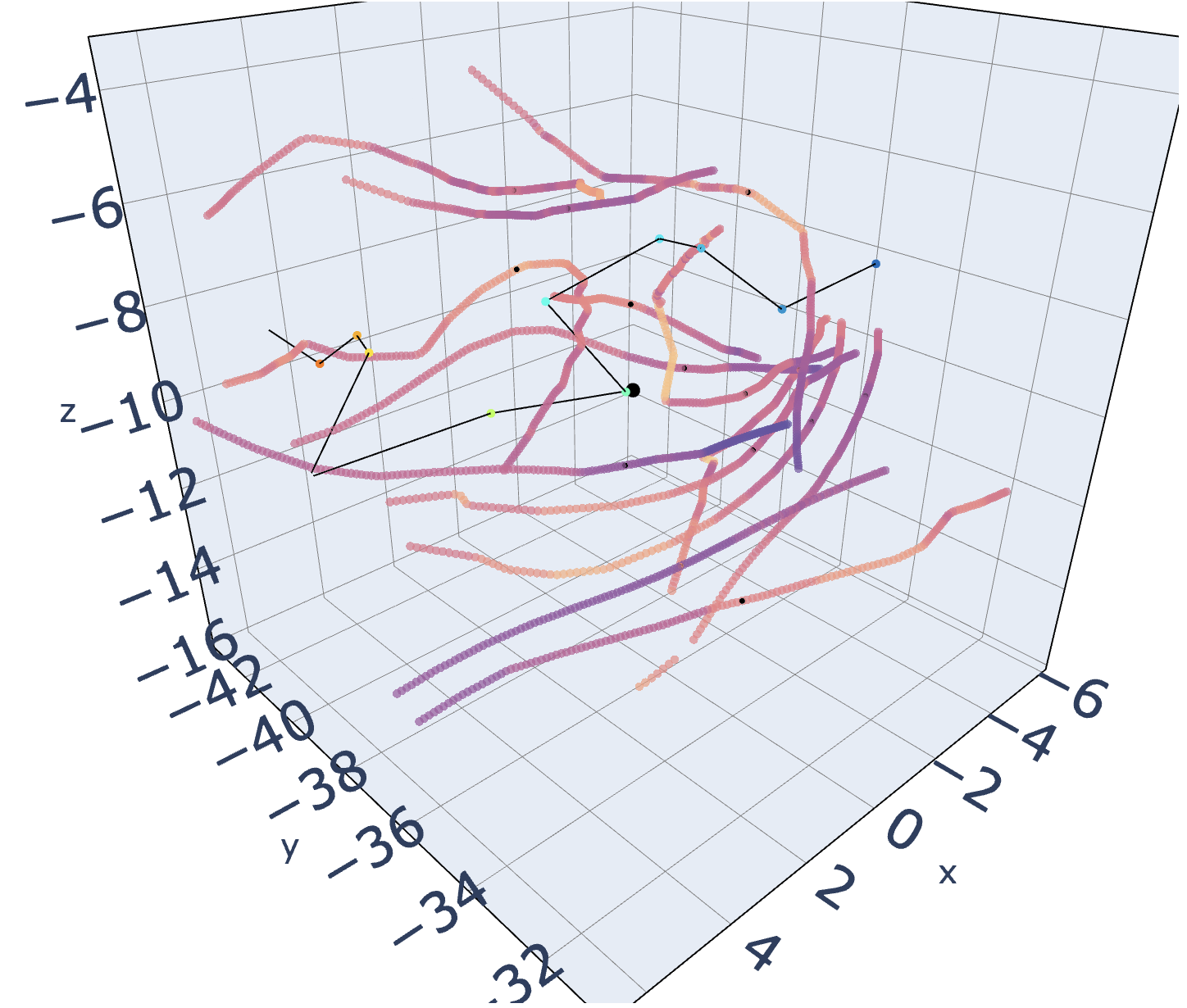}
    \caption{\label{fig_t3} $t = 13815 \; t_g$ }
\end{subfigure}\,
\begin{subfigure}{.04\textwidth}
    \includegraphics[width=\textwidth]{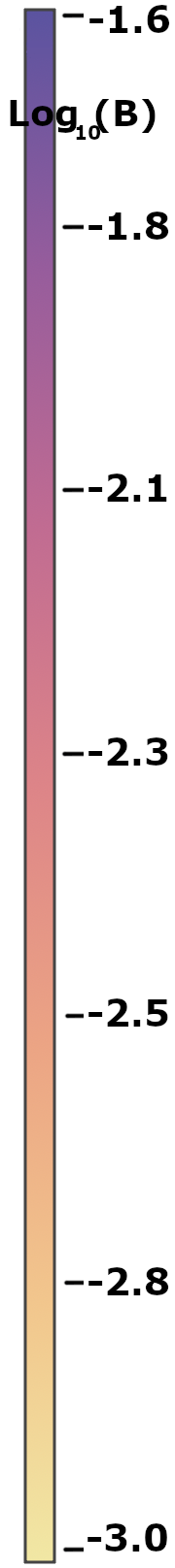}
\end{subfigure}
    \caption{\label{fig_turbulence} Sequence of 3D snapshots of the trajectory of a particle along with magnetic field 
    lines (taken within the black square shown in the left panel of Fig. \ref{fig_inhst_det}).  
    The particle undergoes acceleration by being scattered several times on curved magnetic field lines and strong 
    parallel gradients. 
    The trajectory of the particle is shown as discrete diagnostics colored from blue to red with increasing time connected by a continuous black line 
    (particle enters from the right side of the figures). 
    The central (black) point of the trajectory is the particle's position at the time of the field line snapshot.
    The field lines are colored (blue to red) based on the base $10$-logarithm of the magnitude of the magnetic field.}
\end{figure}

\noindent \textbf{Interaction with structures in the magnetic field}\\
Fig. \ref{fig_turbulence} shows 3D snapshots taken in the neighborhood of a particle's position. The time frame for these snapshots 
is enclosed in the black square in the right panel of Fig. \ref{fig_inhst_det}, 
which is where the particle is gaining energy through several scattering events. 
Fig. \ref{fig_turbulence} (panels \ref{fig_t1} and \ref{fig_t2}) show two consecutive scattering events 
(both scattering events are visible in either panel, the focus is different) occurring in regions where the magnetic field 
magnitude varies over short distances, both in the parallel and perpendicular directions, and, crucially, 
where the magnetic field lines are sharply bent \citep{lemoine2023particle}. 
These two panel also show that structures in the magnetic field are advected without being destroyed by reconnection 
or other processes for significant stretches of time ($\Delta t \geq 20 \; t_g$), 
and still acceleration takes place to a significant degree, even though it requires several scattering events 
to have an appreciable effect on the particle's energy. Panel \ref{fig_t3} shows a mirroring event, 
where the particle interacts with a strong gradient in the direction parallel to the magnetic field, 
thus leading to the textbook behavior due to the conservation of the adiabatic invariants of motion. 
Observing this phenomenon in such a complex scenario also allows to point out that the mirroring alone 
does not have significant impact on the particle energy; it is rather the turbulent environment 
in which the particle is immersed that leads to overall increase in energy.\\ 
Overall, Fig. \ref{fig_turbulence} 
provides a good example of how the interaction of the particles with structure in the magnetic field, and 
not with reconnection events, can lead to energization over several scattering events. This region of the 
domain also likely represents how the bursty behavior in MAD disks is going to affect particle dynamics, 
as it will cause structures in the magnetic field to move in a chaotic manner, enhancing scattering. 
This is however difficult to disentangle from the overall dynamics of the disk itself when it comes to 
the acceleration behavior of particles. 
Furthermore, the time-scales involved in bursts are much longer than those involved in scattering, thus a 
different type of analysis (left for further work) is needed, requiring to combine, statistically, what was 
shown in Sec. \ref{sec_trapping} with the temporal statistics of MAD flows.\\ 
All of this is consistent with the statistical analysis presented in \ref{sec_acceleration}, 
where it was shown that the dominance of negative changes in $< \Delta \Gamma/\Gamma>$ can be explained by 
the fact that most particles, due to their high energies, are predominantly free-streaming or advected 
through the domain. Particle acceleration is observed only in regions where scattering events are expected 
to be more frequent (i.e., where gradients are stronger).\\

\begin{figure}
\centering
\begin{subfigure}{.41\textwidth}
    \includegraphics[width=\textwidth]{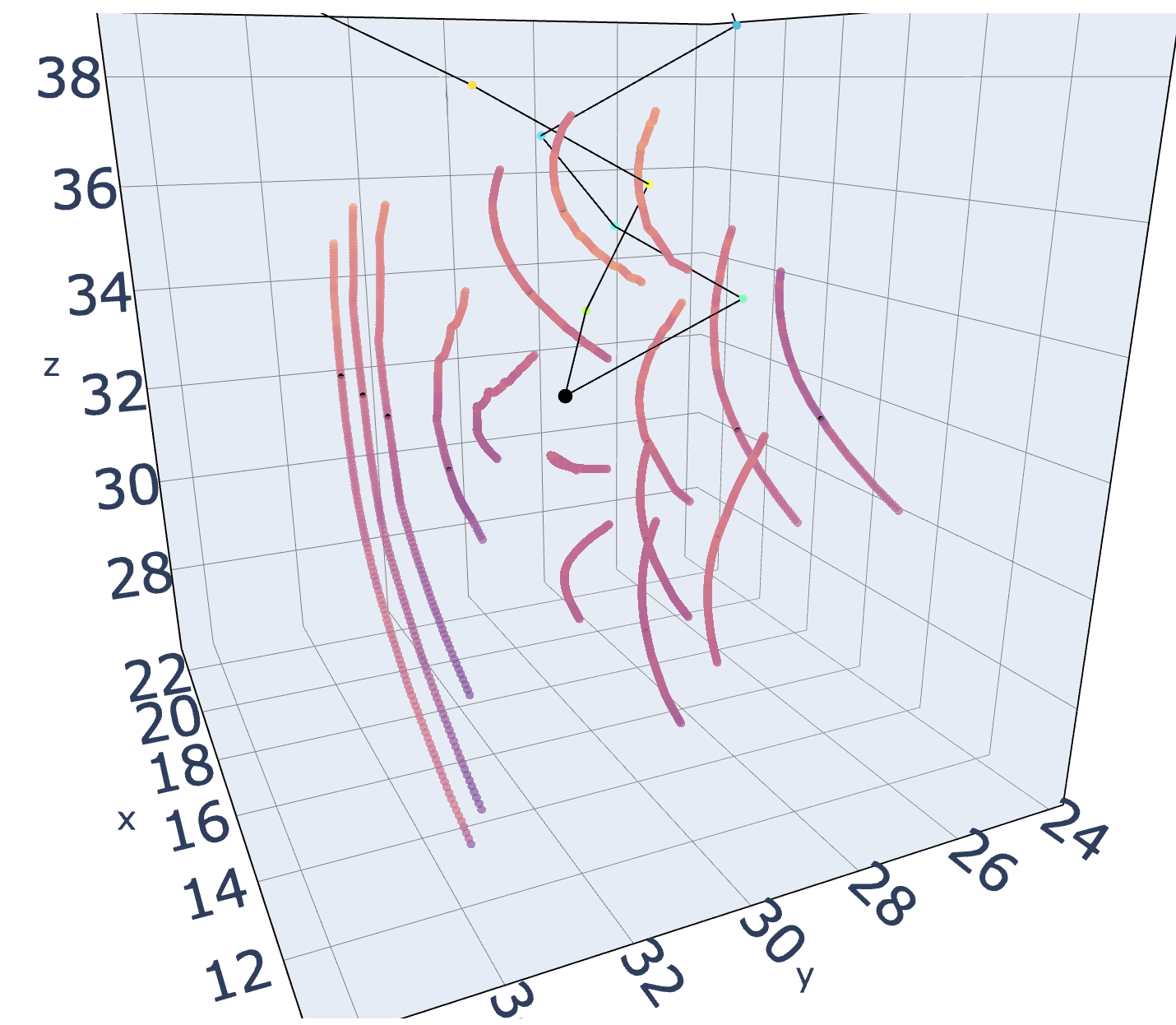}
    \caption{\label{fig_fl1} $t = 13464 \; t_g$ }
\end{subfigure}\;
\begin{subfigure}{.41\textwidth}
    \includegraphics[width=\textwidth]{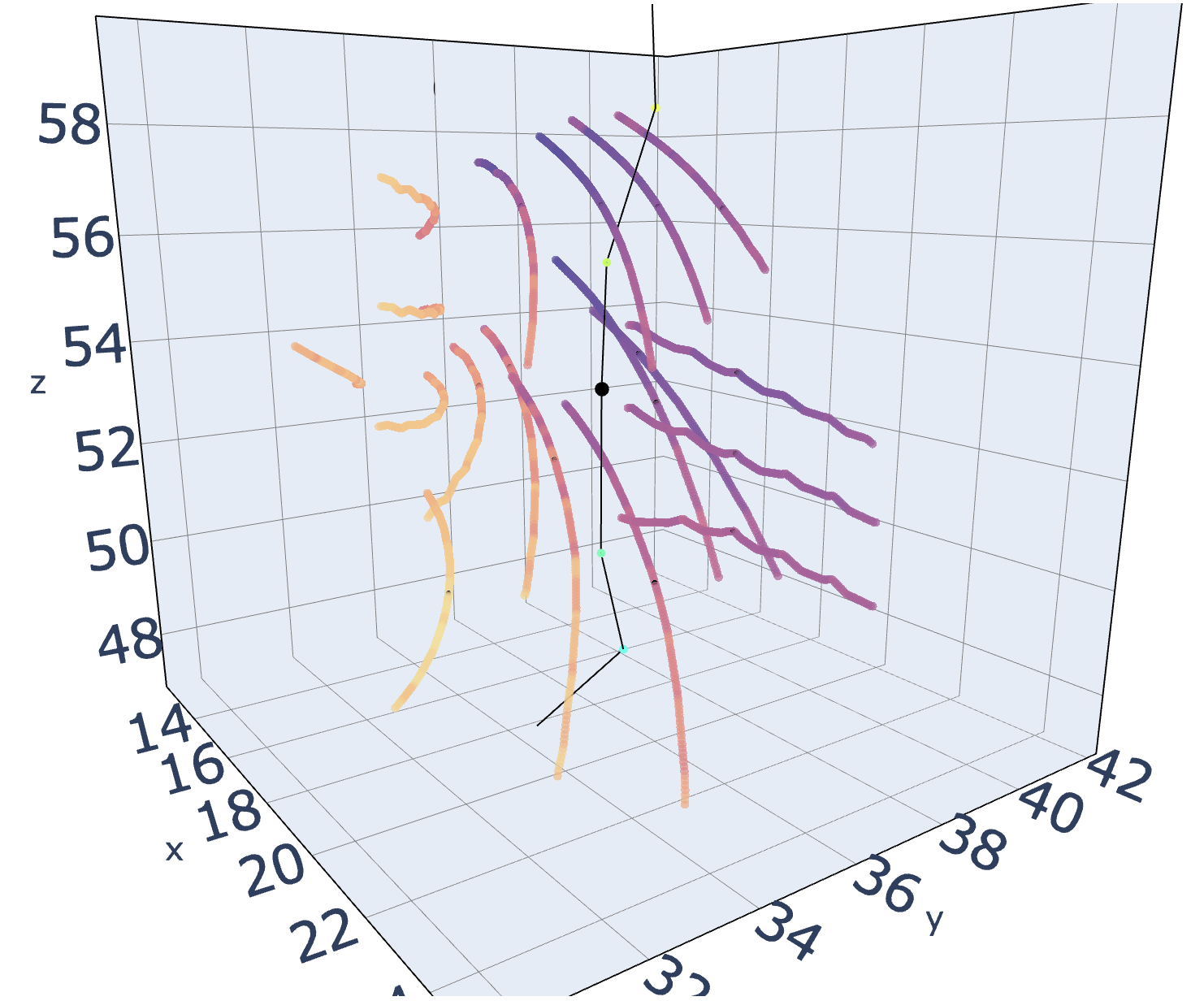}
    \caption{\label{fig_fl2} $t = 13500 \; t_g$ }
\end{subfigure}\,
\begin{subfigure}{.04\textwidth}
    \includegraphics[width=\textwidth]{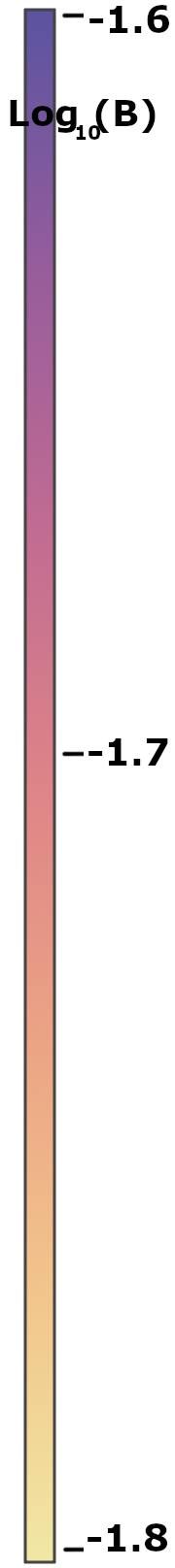}
\end{subfigure}
    \caption{\label{fig_inversion} Sequence of 3D snapshots of the trajectory of a particle along with magnetic 
    field lines (taken within the black ellipse shown in the left panel of Fig. \ref{fig_inhst_det}).  
    The particle undergoes acceleration and inversion of the direction of motion by interacting with the strongly 
    sheared magnetic field. 
    The trajectory of the particle is shown as discrete diagnostics colored from blue to red with increasing time connected by a continuous black line 
    The central (black) point of the trajectory is the particle's position at the time of the field line snapshot. 
    The field lines are colored (bright to dark) based on the logarithm of the magnitude.}
\end{figure}

\noindent \textbf{A novel behavior observed in particles propagating along the jet}\\
A remarkable phenomenon occurs frequently for particles traveling within the funnel region of the MAD simulations.
Some particles propagate for long distances (from $100 - 300 \; r_g$ taking around $100 - 300 \; t_g$, as expected 
for particles with $v \approx c$) along the jet-sheath but then get reflected back towards the BH. This reflection 
is particularly strong for particles that move from the jet spine to the jet-sheath, and thus are affected by a gradient 
that is determined by the numerical floor applied to the density. This is already visible in the trajectories of some 
of the particles chosen in Fig. \ref{fig_inhst_det}. To mitigate this effect, we restrict our analysis to reflections
occurring at $R > 10 \; r_g$, i.e., outside the region most affected by numerical floors (see Fig. \ref{fig_maps_combined}). 
While the numerical strength of the gradient and its localization are affected by the artificial floor, the presence 
of such a gradient is physically expected, thus this effect will occur regardless of the specific floor implementation.  
It should be noted, however, that at these distances from the BH, the simulation resolution is modest. The typical 
Larmor radius for a particle involved in reflections sits at $5 \leq r_L/r_g \leq 9 $, to be compared with the 
resolution at $z \approx 300 \; r_g$ of $\Delta x = 8 \; r_g$, thus the particle dynamics sit just at the limit of 
the resolution, but these reflections occur, occasionally, also in well-resolved regions, as will be shown below.\\ 
The occurrence of these reflections along the jet is not particularly rare, 
showing up as a significant contribution in the average particle distributions.
One can notice the significant tail in the average particle distribution of $\mathbb{M}9$ 
when we extend the range of the energy axis (using logarithmic spacing) in Fig. \ref{fig_avg_dist}.\\
A possible explanation for this reflection can be found in the $\nabla B$ drift. 
A particle traveling in the inner part of the jet is experiencing a helical field with a predominant 
$B_z$ component. It can be accelerated along its motion, and will eventually have a Larmor radius large enough 
to access the region of radial magnetic field gradient between the spine and the sheath. 
The resulting force (drift) from $\bm{F} \sim \bm{B} \times \nabla B$ is along $\hat{\varphi}$, 
and deflects the particle at a given $z$ position in the azimuthal direction, generating a centripetal force 
that puts the particle in an ``orbit'' encircling the jet spine.
If the region of radial gradient of $B$ is sufficiently wide, and the particle has a finite radial velocity, 
this process can occur again, but in the outer layers of the jet the azimuthal component of the magnetic field 
becomes stronger, with the $\nabla B$ drift producing a force directed along $\hat{z}$. 
Depending on the relative direction of the azimuthal and vertical fields, the combination of these two deflections 
can result in an inversion of the direction of motion. If at this stage the particle has left the region of radial 
gradient of $B$, its motion will have effectively been inverted, and the particle will travel towards the BH. \\
In several cases, particles were observed traveling along the funnel region several times (see Fig. 
\ref{fig_inhst_det}, bottom right panel), gaining energy until they finally escaped the region that reflects them back and forth. 
This seems to be a viable mechanism that leads particles that already have very high energies ($\Gamma \approx 
10^8$) to being accelerated by a factor $\times 100$, which would be a remarkable figure if independently 
verified.
Even more interesting is the possibility that lower-energy particles could undergo a similar process, 
as it would provide a very efficient and almost universal mechanism for the acceleration of particles in 
these environments. If, as this analysis supports, turbulent flow-shear acceleration is a fundamental 
process, such particles may undergo acceleration in similar ways, as velocity gradients and shears are 
expected to persist down to scales below the current numerical resolution.
A primary challenge is that, to experience the successive deflections that reverse the direction of motion, 
a particle must already possess a gyroradius large enough to sample the magnetic field gradient. It is also 
interesting to consider that particles that have been accelerated over the length of the jet my be "recycled" 
and contribute to the processes and emission closest to the BH with the energy they have acquired from their 
acceleration along the jet.\\
Some of these particles, when getting close to the BH, have large enough energies that the magnetic field does 
not confine them effectively, and they can get a slingshot assist from the gravitational potential and be 
flung out of the domain. The reason why a similar effect is not observed in $\mathbb{S}9$ is that in this 
simulation the fall-off of the magnetic field along the funnel region is faster than in the MAD cases, 
and thus effects due to magnetic field gradients become quickly less important.\\
Fig. \ref{fig_inversion} depicts one such particle undergoing the process of reflection in a region where the 
grid resolution is still $\Delta x = 0.5$. 
This again underscores that the primary driver of the dynamics, whether it be inversion of the direction of 
motion or acceleration, are structures in the magnetic field related to shear, turbulence, and curvature, 
since in Fig. \ref{fig_inversion} a particle was captured undergoing an inversion in the direction of motion, 
and subsequently acceleration, in a region where the magnetic field does not undergo any reconfiguration.
Rather, the magnetic field presents strong shears that result in forces perpendicular to the gradient itself, pointing in the 
direction the particle ends up moving towards. The intensity of the acceleration increases as the particle 
moves further along the funnel, where the magnetic field is more ordered still. 
The direction of the field lines, that was rapidly and chaotically changing in the case of turbulence 
analyzed in Fig. \ref{fig_turbulence}, is here instead ordered and smooth (neighboring field lines are 
parallel to each other and only change direction gradually). 
This case will resembles more the picture of the particles being accelerated by the shear of the magnetic 
field, which also results in changes in the particle direction, trapping for rather long times, and increases 
in the energy of the particle.\\
We note that the $\Gamma$ trace within the black ellipse in the right panel of Fig. \ref{fig_inhst_det}, 
which corresponds to the snapshots in Fig. \ref{fig_inversion}, appears to exhibit a sudden, nearly 
vertical jump. This is due to the frequency of diagnostic sampling $\Delta t = 3 \; t_g$ not being able 
to capture the rapid dynamics, but the actual time-steps of the simulation are orders of magnitude smaller 
(see Sec. \ref{sec_setup}, and the resolution of the gyration is ensured.\\
As noted in Sec. \ref{sec_escape}, particles initialized in the funnel region often have the largest 
$\tau_{esc}$, but in that section, re-entry in the $R \leq 64 \; r_g$ and $|z| \leq 64 \; r_g$ was not 
considered. When the analysis is expanded to $R \in [0,500] \; r_g$ and $|z| \in [0,500] \; r_g$ 
both $\tau_{esc}$ and $\tau_{in}$ increase significantly. This is not shown due to the decrease in both 
the resolution of the simulation and the statistics of particles this far from the BH. 
However, this behavior now finds a clear explanation in these inversions of the direction of motion, 
that particles initialized in the funnel are more likely to undergo.\\
Finally, Fig. \ref{fig_inhst_det} also shows particles that stayed in the region $r \leq 10 \; r_g$ 
for a few transits, with enough energy to not be captured by the gravitational well. 
It is not uncommon, however, that interaction with the turbulent magnetic field lines, which exhibit
rapidly changing magnitudes and chaotic directions, leads a particle to fall into the BH due to a 
"sudden" change in direction. This, on the other hand, means that particles initially headed towards 
the BH might be redirected, allowing them to avoid infall and resulting in rather long trapping times 
in the innermost region.
While one could expect the process to be isotropic, as the scattering is random in nature, and thus 
roughly half of the particles will be directed towards the BH, while another half will be scattered 
away from it, this may not necessarily be true. 
Firstly, in the vicinity of the BH the turbulence is unlikely to be isotropic because of strong
spacetime curvature and because of the way the dynamics shape the magnetic field.\\ 
A systematic study of turbulence properties in GR environments prospects itself as a natural 
extension of this work, and is currently undergoing. 
Secondly, although the scattering itself might be isotropic, its effect on the average dynamics of 
the particles need not be, due to the imbalance generated by particles falling into the BH. 
Those that get scattered towards the BH end up crossing the event horizon and are thus lost. This 
also explains why closest to the BH the largest average energy increases occur (see Fig. \ref{fig_accel}), 
as particles moving inward have increasing $\Gamma$ from the changes in the metric, but then never 
leave in the opposite direction, leading to an imbalance, as was already noted in 
Sec. \ref{sec_acceleration}.\\

\section{Discussion} \label{sec_discussion}
\subsection{Particle Transport and Acceleration}
After considering the points raised above, we now attempt to assemble a coherent picture of acceleration and confinement 
from these simulations.
When it comes to acceleration, our results in Sec. \ref{sec_acceleration} suggest that scattering, whether off turbulent 
structures or magnetic field curvature, is the primary mechanisms for particle acceleration across all cases analyzed. 
The localization of the acceleration region (see Fig. \ref{fig_accel}) is, in principle, compatible with both the regions 
of intense velocity gradients and turbulent fluctuations (see Figs. \ref{fig_maps_combined} and \ref{fig_spectra}) and 
regions of intense currents (see Fig. \ref{fig_maps_combined}). However, particles at higher energies lose energy (on average) 
by interaction with the disk, while at lower energies there is a more significant energy gain. 
This behavior makes a direct explanation based on the current less likely, since the effect of the electric field 
should be the same (albeit with different magnitudes) regardless of the particle energy for particle that are overall 
moving in the same direction. 
There is also a better match in the amplitude of the fluctuations in regions that provide similar acceleration across 
the MAD simulations (see Fig. \ref{fig_spectra}) than there is for the current, as in Fig. \ref{fig_maps_m99} the current 
is much more localized in the mid-plane than in Fig \ref{fig_maps_m00}, but the acceleration regions in the disk have 
roughly the same extension across the two cases, as shown in Fig. \ref{fig_accel}. Considering these point (and the 
additional arguments presented in Sec. \ref{sec_acceleration} and \ref{sec_individual}) turbulent scattering appears the 
most plausible explanation for the acceleration observed. A contribution by magnetic reconnection cannot be ruled out, 
but our analysis suggests its effect is marginal, as particles seldom intersect a current sheet at the appropriate moment. 
On the other hand, turbulence and curvature are present at all times in regions of strong gradients, and thus provide a 
more robust explanation.\\
One potential observational consequence of the identification of these acceleration regions is a localized enhancement 
of the jet brightness along these regions \citep[e.g.,][]{fromm2022impact}. Particle acceleration in the jet-sheath 
could produce limb-brightening even in the absence of significant Doppler boosting \citep{rieger2004}.
Section~\ref{sec_acceleration} shows that both SANE and MAD disks can accelerate particles. In the MAD runs the acceleration 
is concentrated along the jet, particularly in the sheath, whereas in the SANE runs the efficiency is comparable in the 
disk and the jet, which can be linked to the peculiar structure of the radial flows in these disks (see Fig.~\ref{fig_sigma})
These results support a picture in which turbulent scattering in regions of strong gradients, rather than magnetic reconnection, 
dominates particle energisation.\\
In principle, this kind of study can help to constrain the transport properties in the diffusion–advection equation for 
the phase-space distribution $f(t,x,p)$ of energetic particles with spatial diffusion coefficient $\kappa(x,p)$ and 
source term $S$ \citep[e.g.,][]{Parker1965,Earl1988,Park1995}
\begin{equation} \label{eq_cr_diffusion}
\partial_t f =\partial_x \Big[ \kappa \,\partial_x f \Big] - U \partial_x f
+  \frac{1}{3}\bigg( \frac{dU}{dx} \bigg) \: p \: 
\partial_p f + \frac{\Gamma_s}{p^2} \partial_p \Big[p^4 \kappa \, \partial_p f \Big] +S 
\end{equation} For weakly varying spatial dependencies, for example, this equation approximately describes momentum-space 
diffusion with an added escape term $-f/\tau_{\rm esc}$, representing spatial transport out of the region of interest. Since 
$\tau_{\rm esc}$ cannot be derived from first principles, but influences the spectral shape, an estimate in the innermost 
region is of particular interest. The practical challenge is to achieve sufficient numerical resolution that allows one 
to fully cover the relevant particle-energy range. While the escape times found here are indicative, they are not 
necessarily representative for modeling neutrino fluxes from different AGN \citep{icecube2025}.
Still, it is worth noting that the dominant acceleration mechanism is compatible with turbulent scattering and with 
present turbulent-corona models \citep{padovani2024high,lemoine2025neutrinos,lebihan2026}.\\
It is instructive to extract a mono-energetic particle subset from the simulations (ensuring a sufficiently high
particle count $N \geq 3 \cdot 10^4$ for statistical robustness) to compare the observed dynamics with theoretical
expectations \citep[e.g.,][]{rieger2006} and previous studies \citep[e.g.,][]{kimura2019acceleration}. If particle 
propagation were properly diffusive, the resulting distribution would approach a log-normal form as in simple 
stochastic or shear acceleration models without escape. As a test, we compare the time evolution of the distribution 
to Eq.(14) in \citet{rieger2006}, which captures both the temporal evolution and power-law behavior. While the 
qualitative trend matches the model, with an initial Gaussian-like phase followed by a broadening tail, fitting 
a log-normal distribution proves challenging. The initial phase tends to require extreme parameters ($\xi \geq 6.0$), 
and at later times, the low-energy tail is under-estimated while the high-energy tail is over-estimated. 
The best-fit power-law index is $\xi \approx 1.2$, and no single parameter set fits distributions at multiple 
times. This suggests that the scattering dynamics evolve over time and/or differ from local diffusive (no escape) 
model as considered in \citet{rieger2006}. 
This is not surprising: the simulations are global, include particle loss and potentially capture non-diffusive 
transport, while the reference model is local and assumes conservation of particles. The discrepancy may thus 
reflect the complexity of real accretion flows, where global dynamics, irreversible losses and diverse 
trajectories complicate a universal parameterization.\\
A comparison with previous work offer valuable context, though methodological differences limit direct
quantitative comparison. While the dynamics considered here are fully general relativistic, simulations in
\citet{kimura2019acceleration} neglect gravitational effects on the particles and use Newtonian potentials 
for fluid evolutions. The present study also considers particles with significantly higher energies (and Larmor 
radii), required to resolve gyration even near the BH, where magnetic fields can increase by an order of magnitude 
compared to their values at $r \sim 10 \; r_g$, a predominantly GR effect, particularly prominent in MAD regimes. 
The closest simulation for comparison in this work is $\mathbb{S}9$, that, however, for the high-energy particles 
considered here, exhibits poor confinement, complicating direct comparison. Nevertheless, in reference 
\citet{kimura2019acceleration}, the evolving distributions are fitted with Gaussians, whereas the present analysis 
finds this approach less suitable beyond the initial phases, i.e., well before the $t = 4 \; t_L$ reported in 
Fig. 8 of that work, where $t_L = 2\pi\frac{m}{q}\frac{u^0 c}{B}$).\\
Other relevant studies include \citet{effenberger2018relation}, which explores the relation between scattering, 
crossing, and escape times in idealized isotropic turbulence for various injection schemes, including "beam 
injection," analogous to our flow-aligned setup, and "pancake injection," similar to \citet{kimura2019acceleration}. 
Similarly, \citet{effenberger2018relation} describes super-diffusive transport mediated by Lévy flights, which 
bears resemblance to our observations: particles typically follow weakly-scattering trajectories, dominated 
by advection, but occasionally undergo large deviations, particularly in the funnel region (Fig.~\ref{fig_inhst_det}), 
coinciding with bursts of energy gain. While these studies are local and lack dynamical and general relativistic 
effects, they highlight universal features of particle transport, such as the interplay between coherent motion 
and rare, large-scale scattering events, that are also present in our simulations. This suggests that while 
quantitative comparisons are limited, the underlying physical mechanisms may share commonalities across 
different modeling frameworks.\\

\subsection{Astrophysical context and comparison} \label{sec_realworld}
The simulations presented here probe particles with Larmor radii satisfying $5 \leq \frac{r_L}{r_g} 
\leq 25$ at injection, ensuring adequate orbit resolution throughout the simulation domain up to $r = 
2.5 \; r_g$, where an artificial sink, acting as the horizon, is placed. The chosen range of Larmor 
radii is also a consequence of the desire to have the same energy distribution at all locations where 
injection occurs.\\
In realistic SMBH systems, such as NGC~1068 ($M_{\rm BH} \simeq 
1.6 \times 10^7 M_{\odot}$, $r_g \simeq 2.4 \times 10^{12}$ cm), the particles relevant for the neutrino 
production have $\frac{r_L}{r_g} = \frac{\Gamma m c}{\alpha e B} \ll 1$. The explicit ratio depends 
on the assumed magnetic field strength, which may reach strength up to $B\sim 10^4$ G in the mid-plane 
near the horizon and drop to $B \sim (10^2-10^3)$ G in the corona and farther out 
\citep[e.g.,][]{yuan2014hot,inoue2020,lemoine2025neutrinos}. The proton energy required to produce the 
IceCube neutrinos \citep{icecube2022evidence} is $E_p \sim 0.1$ PeV ($\Gamma \sim 10^5$), corresponding 
to Larmor radii $r_L \sim 3\times 10^{11}/B$ cm, i.e., a ratio $r_L/r_g \lesssim 0.1/B$.   
Resolving such a small ratio for, e.g. $B>100$ G, would require a grid spacing $\Delta x < 10^{-3}\,
r_{g}$, which is difficult to achieve with current numerical capabilities. An additional limitation
arises from the fact that ADAF-like flows simulated with MAD and SANE tori are not expected to remain 
physically self-consistent at the high accretion rates inferred for NGC 1068, where radiative cooling 
can no longer be neglected. In terms of the normalized AthenaK units, the Eddington accretion rate for
the latter corresponds to $\dot{m}\sim 5 \times 10^5$, several orders of magnitude above the rates 
simulated here (see Fig.~\ref{fig_history}).\\
Nevertheless, we can attempt an extrapolation that might prove informative, highlighting the relevance
of the cases considered here relative to known sources, despite the noted limitations. 
NGC 1068 is not a strongly jetted galaxy, and instead thought to host a compact corona-like 
region \citep{padovani2024high}. Thus the case that seems most relevant among those analyzed 
in this paper is the $\mathbb{S}9$ simulation, that also has a particle energy range somewhat 
closer to the mentioned $\Gamma \sim 10^5$, though still significantly higher. However, as 
discussed in Sec. \ref{sec_escape}), SANE disks provide only limited confinement, as the 
magnetic field strength decreases rapidly with distance from the black hole. This reflects
the absence of strongly magnetized regions beyond the innermost ($r \leq 20 \; r_g$) region 
of the disk. Strongly magnetized regions with strong gradients also appear, from the 
simulations shown, to be much more efficient at accelerating particles, which may in turn
enhance neutrino production in hadronic scenarios. This leaves open the possibility that 
particles with lower energies than those considered here could be confined by the disk for
longer times, since their smaller gyro-radius could prevent (or at least delay) their access 
to regions of significantly lower magnetic field. This would allow more time for interactions 
with the fluid, potentially leading to interactions similar to those observed in the MAD cases. 
However, verifying such effects requires simulations with higher dynamic range than employed 
here, and thus the true confinement properties of SANE disks at lower particle energies remain 
somewhat uncertain.\\
For efficient neutrino production the disk must contain an extended region of strong, 
preferably turbulent, magnetic field. Our MAD simulations provide such conditions, but 
they not reproduce the assumed properties of NGC 1068. None of our simulations reveal 
a clear, magnetized corona that is believed to be one of the defining features of neutrino 
emitters like NGC1068 \citep{padovani2024high}. We remind the reader that, in this paper, 
we adopt a loose definition of ``corona'' as a region of strongly magnetized plasma that 
is gravitationally bound (as determined by, e.g., the Bernoulli parameter criterion) 
and lacks wind-like outflows \citep{hawley2004general}.
Features resembling coronas have, however, been observed in recent simulations that 
include radiative transport and achieve higher accretion rates 
\citep{zhang2025radiation}.\\
That said, extrapolating our results to lower particle energies suggests an interesting 
scenario. The magnetic field in the SANE case may be sufficient to confine lower-energy 
particles long enough for efficient acceleration, while particles above a certain energy
threshold escape with minimal deflections. This could render SANE disks more 
effective at letting cosmic-ray particles escape.
Notably, a relatively hard ($\xi \sim 1-2$) low-energy proton spectrum with a very 
soft high-energy tail appears needed to model the neutrino emission in NGC~1068 
\citep[e.g.,][]{murase2022hidden,lemoine2025neutrinos,lebihan2026}. 
Figure~\ref{fig_avg_dist} suggests that, provided particle feedback on the turbulence 
can be neglected, efficient acceleration of low-energy particles can yield a hard 
particle spectrum, while subsequent efficient escape at higher energies may lead 
to strong spectral softening.

\section{Conclusions} \label{sec_conclusion}
In this work, we have studied the dynamics of energetic particles in advection-dominated accretion 
flows (ADAF), observing acceleration factors up to $10^2$ in funnel regions through interaction 
with turbulence. This acceleration is likely driven by shear-type mechanisms \citep{rieger2019}, 
or more generally, by interactions with flow-driven magnetic field structures 
\citep[e.g.,][]{lemoine2019generalized, lemoine2025effective}, such as turbulent, curved, or 
sheared field lines. Confinement times of particles in the wider disk region can significantly 
deviate from diffusive and free-streaming estimates (see Tabs. \ref{table_escape} and 
\ref{table_estimates}), sometimes by up to an order of magnitude. We also identify new 
dynamics specific to the global, GR nature of our simulations.
In particular, the global nature of the background GRMHD simulations allows to study particle 
propagation across regions with significantly varying parameters, that are self-consistently 
connected by the fluid simulations. This reveals dynamics with features reminiscent of
L\'evy flights, with small, random scattering events that confine particles in limited regions, 
occasionally interrupted by large, sudden jumps in both energy and location.
The observed, complex particle dynamics 
at higher energies indicate that using spatially-constant (energy) diffusion coefficients 
or advection dynamics to model particle transport may not be valid beyond the low-energy
regime and thus lead to notable discrepancies.
The (GR) nature of the simulations also enables the study of black hole spin effects 
on particle dynamics in ways that have no Newtonian equivalent. In particular, the 
presence of strongly magnetized outflows and funnel regions \citep{blandford1977electromagnetic,
blandford1982hydromagnetic} allowed us to observe, for the first time, a reflection process 
that particles can undergo during propagation along the jets formed in the high-spin MAD case. 
Since the most intense acceleration region occurs in the funnel-sheath region, these 
reflections allow particle energization to last longer than would otherwise be estimated.\\ 
Overall, the picture that emerges is that strongly sheared velocity gradients drive turbulence, 
which in turn generates and sustains the magnetic field structures that interact with particles. 
This is particularly important because such dynamics can cascade self-similarly across the inertial 
range of the turbulent spectrum, implying that the same acceleration mechanism operates across a 
wide range of scales, including for lower-energy particles, that are of particular interest for
astrophysical applications. Indeed, these observations prove relevant for understanding the origin 
of  ultra-high-energy cosmic rays in AGN jets \citep[e.g.,][]{rieger2022,wang2024}, particularly 
if particles with lower injection energies also undergo similar processes. 
This is likely the result of two compounding processes. When the gyro-motion of the particle 
occurs on scales much smaller than the velocity gradient, the particle interacts with the 
small-scale turbulent structures generated by the cascade triggered by the large-scale velocity 
gradient. Once the gyro-radius is large enough, the particle samples the velocity shear directly, 
enabling efficient acceleration. 
Tentative extrapolations of these results to neutrino-emitting Seyfert galaxies such
as NGC~1068 suggest that efficient acceleration of low-energy particles in the accretion 
flow near the black hole can yield a hard particle spectrum, while subsequent efficient 
escape at higher energies may lead to strong spectral softening. It must be stressed, however, 
that particles in our simulations are purely passive tracers and do not feedback on the fluid.\\
The results presented here underscore the need for future developments bridging the large gap
between particles involved in neutrino production or non-thermal emission and the scales of 
the system of interest. Resolving the Larmor orbits of such particles is computationally 
prohibitive, owing both to the spatial resolution required and to the high Larmor frequencies 
involved. A possible way to overcome these limitations is to extend AthenaK to evolve particles 
within the gyro-center approximation \citep{northrop1961guiding,bacchini2019generalized,zou2024new}. 
While this approach would enable studies of particle interactions with the large-scale structures 
resolved by the flow, capturing the transport and acceleration of low-energy particles may require 
advanced subgrid models that account for interactions with plasma structures unresolved by the 
GRMHD simulation. Incorporating radiative cooling, or even radiative transport 
\citep{jiang2019global, zhang2025radiation}, will be important to accurately model the flow 
dynamics and corona formation in black hole systems at high accretion rates. 
Similarly, exploring additional flow configurations, such as ADIOS or other RIAFs, will be 
of interest for understanding a broader range or accretion \citep[e.g.,][]{blandford1999fate, 
begelman2012radiatively, abramowicz2013foundations}. 
Extending the dynamic range of the fluid simulation, currently underway, 
will offer new insights, while including magnetic reconnection \citep{ripperda2022black}, 
retrograde and tilted tori \citep{mckinney2013alignment, das2024three}, and turbulence-driven 
acceleration mechanisms will further deepen our understanding of particle energization. 
Lastly, including more physics to model the GRMHD flows, other than the already mentioned 
radiative transport, such as resistive effects \citep{ripperda2022black} or even extending the 
approach to two-fluid models \citep{most2022modelling}, would represent important developments 
to the modeling, as the former improves the fidelity of the magnetic field modeling by explicitly 
having physically motivated reconnection mechanisms, rather than the grid-scale dissipation 
of the ideal models, and the latter would allow for different turbulent modes to develop in 
the plasma, thus potentially altering the interaction between particles and fluid.

\acknowledgements
The authors would like to thank A. Bohdan, S. Cerri, M. Lemoine, S. Aerdker and J. Wang for 
interesting discussions on the topic. The anonymous referee is also thanked for their 
constructive and detailed feedback. This work was partially funded by the Deutsche 
Forschungsgemeinschaft (DFG, German Research Foundation) under Germany´s Excellence Strategy 
– EXC 2094 – 390783311. 

\bibliographystyle{unsrtnat}
\bibliography{./Main_Bib}

\appendix 
\section{Validation of the particle pushers} \label{sec_validation}

\begin{figure}
\centering
    \includegraphics[width=\DefFigWidth]{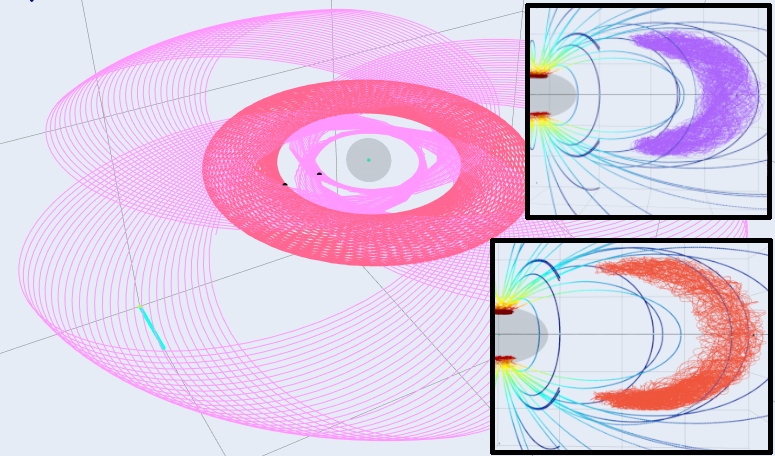}
    \caption{\label{fig_en_cons_num} Validation of the particle pusher used in this study done by 
    verifying that particles remain on stable orbits in regions of strong spacetime curvature. 
    The main figure shows two particles moving in the mid-plane of a system with a Kerr BH ($a=0.9$) 
    at its center. The orbits precess and cover large radial extents (from $r \approx 5\; r_g$ up 
    to $r \approx 80 \; r_g$, but the particles neither fall into the BH nor escape at infinity 
    even after $t \geq 10^4 \; t_g$. The insets show the projection in the poloidal plane of the 
    chaotic particle trajectories that result from the setup of a static dipole in the Kerr metric.
    Colored lines represent magnetic field lines, with color indicating the log of the magnetic
    field magnitude.
    The purple trace corresponds to a metric with $a=0.9$, a particle with energy $E = 0.885$ and 
    angular momentum $L=-7$ initialized at $r_0 = 10.5 \; r_g$. The red trace corresponds to a 
    metric with $a=0.6$, a particle with energy $E=0.9$, and the same $L$ and $r_0$ as the previous 
    one. Crucially, the bounce points remain stable over time, and the radial extent of the orbit 
    remains confined even in such extreme conditions and for $t \geq 2.5 \cdot 10^4 \; t_g$.}
\end{figure}

Similarly to the tests done in \citet{bacchini2018generalized, bacchini2019generalized}, the geodesic terms 
in the implicit pusher can be tested by initializing particles with values of the angular momentum that 
result in stable (precessing) orbits \citep{levin2008periodic, grossman2012harmonic} when positioned close 
to the BH (not necessarily only on the mid-plane). In Fig.~\ref{fig_en_cons_num} two such orbits are shown, 
integrated for $t = 10^4 \; t_g$, demonstrating the ability of the pusher to conserve the energy of the 
particles.
Similarly, the full particle push, including electromagnetic terms, can be tested for stability by setting 
up a static dipole field in the Kerr metric \citep{takahashi2009chaotic} and verifying that the bounce positions 
of the particles initialized with appropriate energies and angular momentum do not change over time. 
Figure~\ref{fig_en_cons_num} shows the projection of the particle trajectories that undergo such repeated 
bounces in one poloidal plane, and one can notice how the bounce points do not change over time
(this test was run for $2.5 \cdot 10^4 \; t_g$) and the particle trajectory is well confined radially. 
The precision of this pusher depends mostly on the time-step used to evolve the simulation and on the tolerance used 
for the iterative procedure.\\

\section{Impact of particles on code performance} \label{sec_performance}
\begin{figure}
\centering
    \includegraphics[width=\DefFigWidth, trim=0 0 7em 10em, clip]{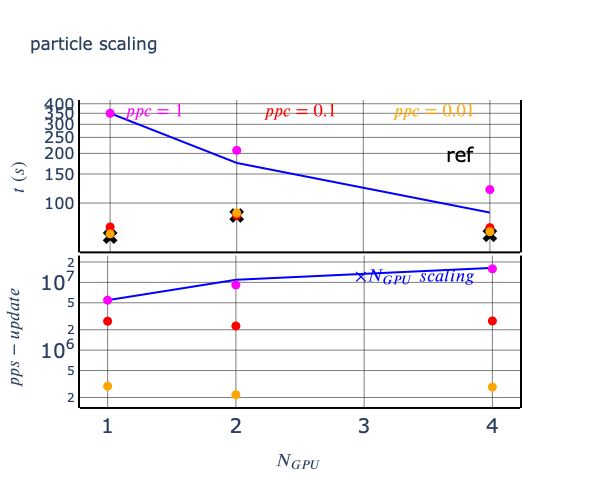}
    \caption{\label{fig_performance}Summary of performance impact of particles on simulations performed 
    with AthenaK. Points of different colors indicate different particle number 
    (particle per cell ``ppc'') as a function of the varying number of GPUs (NVIDIA A100) employed. 
    The top plot shows the total time taken by simulations. The black crosses show the reference case 
    with no particles and only GRMHD computations. The blue line shows the ideal scaling where duration 
    of a simulation scales exactly like $1/N_{GPU}$.
    The bottom plot shows the number of particle per cpu-second (``pps'') updates. Here the blue line 
    shows the performance scaling expected expected naively by scaling like $\times N_{GPU}$.}
\end{figure}

Regarding the specifics of the implementation in AthenaK \citep{stone2024athenak}, particles are handled at 
the level of the meshblockpack, as they can easily move across regions and the meshblockpack contains all 
the information about the fields needed to carry out the particle push.\\
The average number of iterations in realistic applications, where electromagnetic fields are present and 
turbulent, is usually around $10$, though this also depends on the chosen tolerance, and some particles 
can take up to $50$ iterations. 
The rate of failure for the convergence of the iterative method is very small, roughly $1$ particle every 
$2\cdot 10^5$ fails to converge within the upper limit of iterations (set at $200$). In this case the 
fallback strategy is an explicit mid-point update. The failure rate improves significantly over time (to 
fewer than $1$ particle in $1\cdot 10^6$ in the later stages) as particles leave the most challenging 
region of the system closest to the BH.\\
We run sets of simulations using $0.01$, $0.1$ and $1$ particles per cell (``$ppc$''), varying from $1$ to 
$4$ GPUs (NVIDIA A100 GPUs) and compare them also with a simulation running exclusively the GRMHD 
computations (the ``$ref$'' points in Fig. \ref{fig_performance}).
The particle parameters are chosen such that the time-step is not further constrained by particle motion. 
Since the particle solver is iterative, the performance impact presented here is only indicative, and 
challenging systems might require more average iteration numbers than presented here. 
The simulations are set up in such a way that each meshblockpack contains the same number of particles. 
Indeed, load balancing is an outstanding issue in the particle implementation of AthenaK, currently being 
worked on.\\
We note that there is a slight degradation in performance going from $N_{GPU} = 1$ to $N_{GPU} = 2$ due to 
the introduction of communication in the execution of the program and the small size of the mesh chosen as 
reference. In other words, at low particle numbers, the reference problem we chose for this indicative 
example of performance scaling is not ideal for multi-gpu execution as it is communication/memory-bound.
This is particularly visible in the bottom panel of Fig. \ref{fig_performance}, where for $ppc < 1$ the 
particle-per-cpu-second (``$pps$'') updates are degrading with $N_{GPU}$, due to the inefficiency of the 
communication, while at $ppc = 1$ the problem becomes computationally bound, and the expected scaling is 
recovered, with communication playing a minor role.
The top panel of Fig. \ref{fig_performance} shows that at lower particle densities the impact on performance 
is negligible, and thus the more in-depth scaling analysis presented in \citet{stone2024athenak} applies, but 
at higher particle density the cost of particles can quickly become dominant.
Indeed, at $ppc = 1$ one can notice how the scalings with $N_{GPU}$ follows very closely the expected 
$1/N_{GPU}$ scaling, with minor inefficiencies introduced by communication. Thus $ppc = 1$ makes the 
problem computationally bound no matter the size of the original GRMHD problem.\\
Aside from the increased computational cost of the simulations, there is the additional cost of the 
communication of the arrays that contain particle information, that comes with an additional memory 
demand. Since individual particles are tracked, at each timestep, the code must check in each meshblockpack 
which particles has crossed the boundaries and either send it to the neighboring region of the domain 
or destroy it if it has left the domain altogether. Thus there are two additional arrays that need to 
be allocated to contain the information of these particles. In principle, though this is an extreme 
corner case, all particles within a rank might be sent at the same time, or all might be destroyed, 
and there is no way to know this a priori, thus the arrays have to be allocated with sufficient size 
to accommodate for this eventuality. Concretely, allocating arrays with $1/4$ of the total size of the 
total array is sufficient for the present usecases.\\
Load balancing can be straightforwardly adapted to consider particles by assigning a default weight 
to each particle, such that meshblocks with more particles might be better distributed across computational 
resources.

\section{Coarse-graining Analysis} \label{sec_coarse_graining}
\begin{figure}
\centering
\begin{subfigure}{\DefFigWidth}
    \includegraphics[width=\textwidth, trim = 0 0 0 10em, clip]{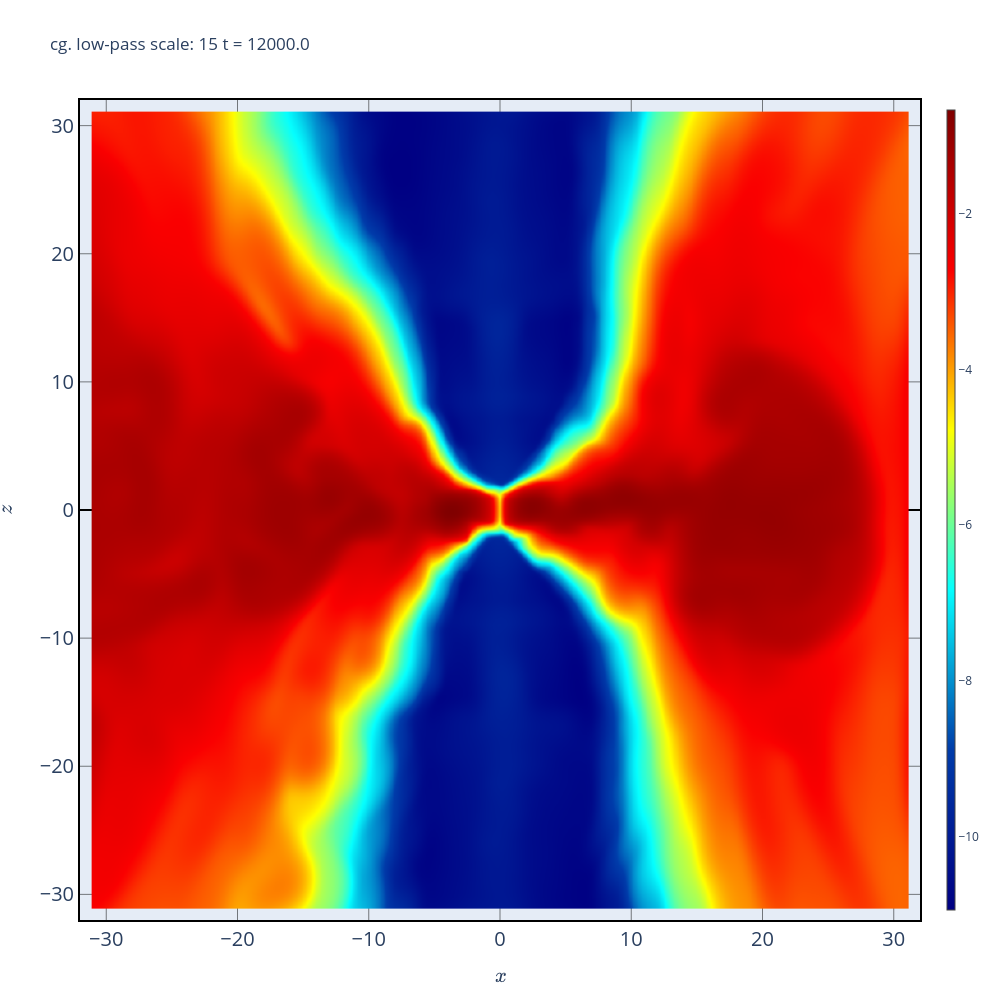}
    \caption{$\mathbb{M}9$. Low-pass filter ($\ell = 15$).}
\end{subfigure}
\begin{subfigure}{\DefFigWidth}
    \includegraphics[width=\textwidth, trim = 0 0 0 10em, clip]{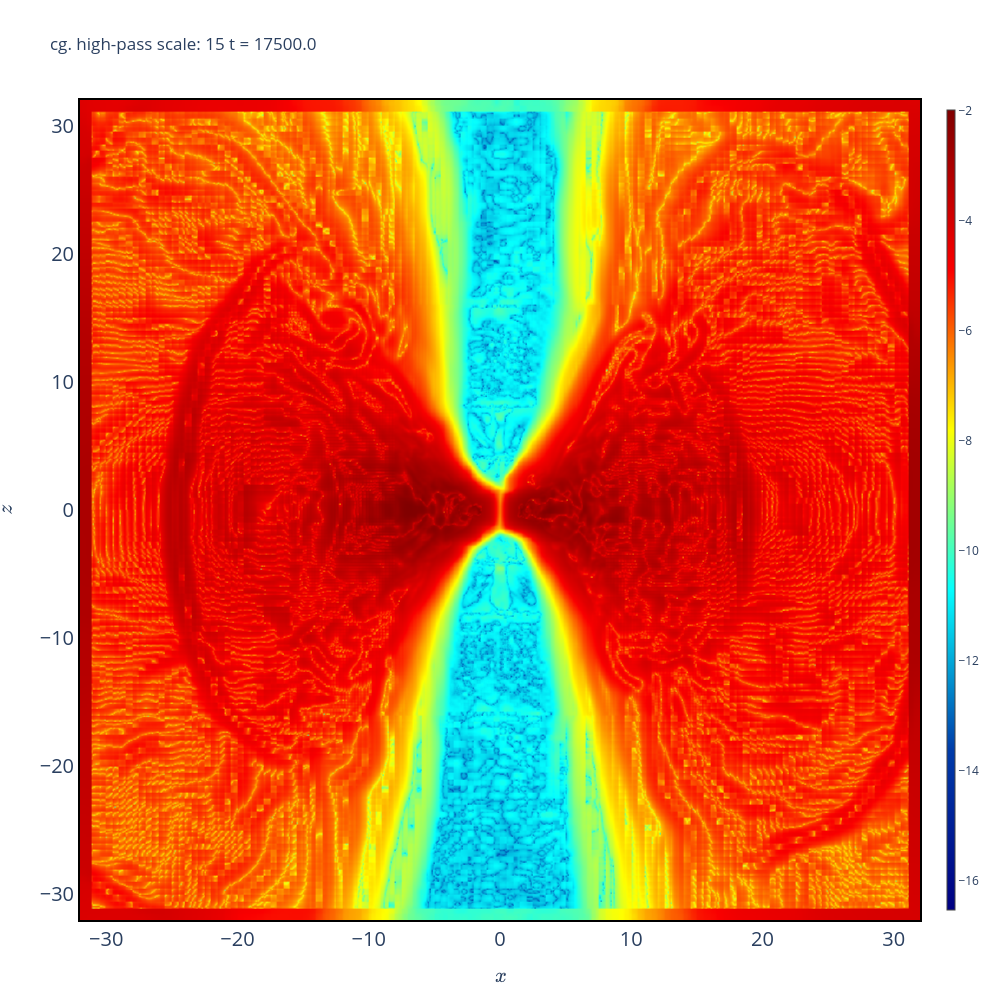}
    \caption{$\mathbb{S}9$ High-pass filter ($\ell = 15$).}
\end{subfigure}
    \caption{\label{fig_filtered} Heatmaps showing the result of the coarse-graining procedure on the density field.}
\end{figure}

\begin{figure}
\centering
\begin{subfigure}{.31\textwidth}
    \includegraphics[width=\textwidth, trim = 0 0 0 10em, clip]{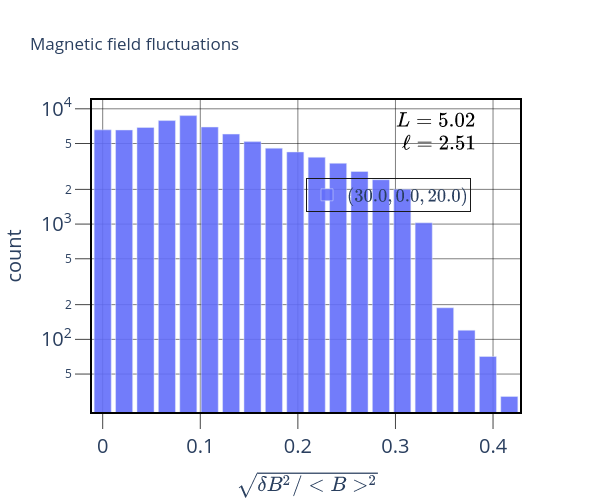}
    \caption{Location $(30.0,0.0,20.0)$.}
\end{subfigure}\,
\begin{subfigure}{.31\textwidth}
    \includegraphics[width=\textwidth, trim = 0 0 0 10em, clip]{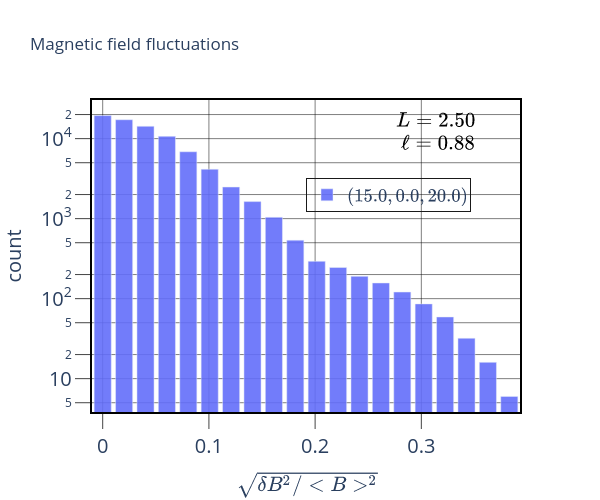}
    \caption{Location $(15.0,0.0,20.0)$.}
\end{subfigure}\,
\begin{subfigure}{.31\textwidth}
    \includegraphics[width=\textwidth, trim = 0 0 0 10em, clip]{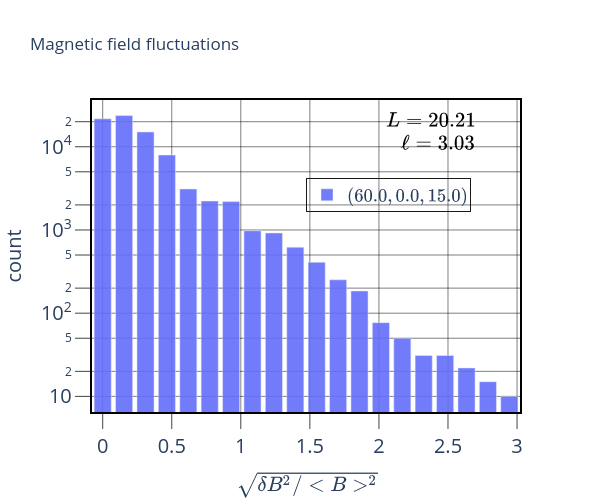}
    \caption{Location $(60.0,0.0,15.0)$.}
\end{subfigure}
    \caption{\label{fig_fluc} Distribution of the magnetic field amplitude fluctuations at locations of interest for the analysis in the main body of the paper from simulation $\mathbb{M}9$.}
\end{figure}

Coarse-graining is used as a technique to analyze the turbulent properties of a medium when the conditions for the application of Fourier analysis (homogeneous background, periodic directions etc...) are not satisfied. In this appendix we describe the implementation of the coarse-graining process as applied in this paper.\\
As a preliminary step, we construct, starting from the data stored on the multi-refinement mesh, an array with uniform resolution, by taking the highest refinement level available and applying it to the whole domain $[-32,32] \times [-32,32] \times [-32,32] \; r_g$ (larger domains can run into memory issues on some smaller machines, often requiring more than $64 \, GB$), and filling in the "newly created" points when refining cells by performing tri-linear interpolation of the values in the cell itself and its neighboring ones in post-processing. This concretely means taking the most refined cell in the simulation data as the reference scale $dx$, and defining, at each point of the domain, a relative scale $\ell_n = dx_n/dx$. Then the value of the current cell is placed at coordinate $x=0$ of a cell that covers the domain $[-0.5 \, dx_n , 0.5 \, dx_n]$, and the values from the neighboring cells will be placed at positions $x_l = -0.5\cdot(1+dx_m)$ and $x_r = 0.5\cdot(1+dx_p)$, where $p,m \in [n-1,n+1]$, as the neighboring cell could be in a region with different refinement. These three values are then used to interpolate the values $\ell_n$ cell created from the refinement, that will be located at $x = ( -0.5 + i/(\ell_n - 1 ) ) \,dx_n$ for $i \in [0, \ell)$. Currently this approach does not include values from the cells neighboring any given one at the corners or edges, only through faces.\\
The uniform array thus created is used to perform coarse-graining (aka filtering) analysis \citep{zhao2025calculating}. 
In short, this consists in defining a coarse grained field $\bar{u}_\ell$ at scale $\ell$ by convolving 
the original field $u$ with a scaled kernel $G_\ell (\bm{r}) = \ell^{-n} \; G(\bm{r}/\ell) $, 
that should satisfy the condition of being third-rank:
\begin{equation}
\bar{u}_\ell(x_1) = \int d x_2 \; G_\ell(x_1 - x_2) \; u (x_2) \qquad G_\ell(x_2) = c\;g_\ell(x_2) - c' \; g_{\ell'} (x_2 - x_0) - c' \; g_{\ell'} (x_2 + x_0)
\end{equation}
where $g_\ell (x)$ is a scaled Gaussian kernel. Since the system we're considering is three-dimensional, we consider the product of 3 1-dimensional kernels to perform the convolution.\\
When doing the convolution, we consider points up to a distance $\vert \bm{x} - \bm{r} \vert \leq 2 \ell$ to ensure that a large enough area around each point is used for the convolution (i.e. to limit "boundary effects" from the kernel being applied to discrete points).\\
From this coarse-grained (low-pass filtered) field, one can reconstruct a fluctuation spectrum that can be shown \citep{zhao2025calculating} to be equivalent to the usual Fourier spectrum in situations where both can be applied:
\begin{equation}
\bar{E}(\kappa_\ell) \equiv \frac{d}{d \kappa_\ell} \left\langle \frac{1}{2}\vert \bar{u}_\ell (\bm{x}) \vert^2 \right\rangle_X = - \frac{\ell^2}{L} \frac{d}{d\ell} \left\langle \frac{1}{2}\vert \bar{u}_\ell (\bm{x}) \vert^2 \right\rangle_X
\end{equation}
as the domain for the averaging we choose $\vert x - x_0 \vert \leq \ell/2$. After the scale of the convolution domain crosses $40$ points (in all directions, since we're doing the convolution isotropically), the domain of the convolution is incremented by $2$ points at each further step. This is mostly to save on post-processing time while only affecting a region of the spectrum that in logarithmic plots already appears very compressed.\\
Furthermore, it has been shown \citep{zhao2025calculating} that using the high-pass-filtered spectrum allows to replicate spectra with steeper index $\xi > 2p + 3$ (compared to $\xi > p + 2$ for low-pass filter), 
where $p$ is the rank of the kernel used to perform the coarse-graining. 
The high-pass-filtered spectrum is obtained from the coarse-grained field simply by replacing $\bar{u}$ in the previous equations with $u' = u -\bar{u}$. 
Examples of the low-pass and high-pass filtered fields are shown in Fig. \ref{fig_filtered}. In these figures one can also notice that the outermost regions of the domain don't undergo the coarse-graining procedure. 
This is done to simplify the operations on the multi-level grid, avoiding having to look for neighboring meshblocks outside the area of interest, since in any case the analysis is carried out in the innermost regions of the coarse-grained domain.\\
The coarse-graining procedure also allows to estimate the magnitude of the magnetic field fluctuations locally, as shown in Fig. \ref{fig_fluc} for two different locations of interest for the analysis in the main paper. 
The average value of the magnetic field is computed within a certain sub-domain over which then the coarse-graining procedure is performed, and from the high-pass filter spectrum one can derive the distribution of the fluctuations.
As visible, in the jet-sheath region, where the magnetic field is strong and with a stable structure, the fluctuations are smaller when compared to the disk/outflow region, where turbulence is more predominant and doesn't allow the formation of stable structures.
In the disk, in fact, most of the fluctuations have amplitude comparable to the average magnetic field, meaning that the region is extremely turbulent.
Getting closer to the BH (not shown), reduces the amplitude of the fluctuations, as the magnetic field is strengthened and its structure more strongly defined by interaction with the boundary conditions of the BH.

\section{Table of symbols}

\begin{table}
\begin{ruledtabular} 
\begin{tabular}{|l|c|c|c|} 
\textbf{Quantity} & \textbf{Symbol} & \textbf{Definition} & \textbf{Use}\\ \hline \ 
Lapse function & $\alpha$ & $\sqrt{-\frac{1}{g^{00}}}$ & Time dilation between observers. \\ \hline \ 
Shift vector & $\beta^i$ & $-g^{0i}/g^{00}$ & Spatial shift in 3+1 decomposition. \\ \hline \ 
Spacetime metric & $g_{\mu \nu}$ & -- & Full metric tensor. \\ \hline \
Spatial metric & $\gamma_{i j}$ & Spatial part of $g_{\mu \nu}$ & Defines spatial distances in 3+1. \\ \hline \ 
FIDO 4$-$velocity & $n_{\mu \; FD}$, $n^{\mu \; FD}$ & $(-\alpha,0,0,0)$, $(1,-\beta^i)/\alpha$ & 4$-$velocity of fiducial observer. \\ \hline \ 
Lorentz factor (FIDO frame) & $\Gamma$ & $\sqrt{1 + \gamma^{ij}u_iu_j}$ & Measured by fiducial observer. \\ \hline \
Time component of particle 4$-$velocity & $u^0$ & $\Gamma/\alpha$ & In coordinate basis \\ \hline \ 
Particle 4$-$velocity & $u_\mu$, $u^\mu$ & -- & Covariant/contravariant components. \\ \hline \ 
Coordinate 3$-$velocity & $v^i$ & $d x^i/dt$ & Spatial velocity measured at infinity. \\ \hline \
Proper time & $\tau$ & -- & Particle's proper time. \\ \hline \
Rest mass density & $\rho$ & -- & Proper rest mass density of fluid. \\ \hline \
Fluid 4$-$velocity & $w^\mu$ & -- & Fluid 4$-$velocity. \\ \hline \
Magnetic field (coordinate basis) & $B^i$ & -- & Eulerian magnetic field. \\ \hline \
Magnetic field (fluid rest frame) & $b^\mu$ & ${}^*\!F^{\mu \nu} w_\nu$ & Magnetic field in fluid rest frame. \\ \hline \
Magnetization & $\sigma$ & $b^\mu b_\mu/(\rho c^2)$  & Magnetic over rest$-$mass energy density. \\ \hline \
Internal energy & $\epsilon$ & -- & Internal energy density. \\ \hline \
Adiabatic index & $\chi$ & $4/3$ & Ratio of specific heats for relativistic gas. \\ \hline \ 
Enthalpy & $h$ & $\rho + \chi \epsilon + (\alpha B)^2/2$ & Relativistic enthalpy. \\ \hline \
Alfv\'en velocity & $v_A$ & $\frac{\alpha B}{\sqrt{h}}$ & Characteristic speed of magnetic waves. \\ \hline \ 
Bernoulli parameter & $Be$ & $ -\frac{u_0 h}{\rho} - 1$ & Measure of fluid binding energy. \\ \hline \ 
Electromagnetic tensor & $F^{\mu\nu}$ & -- & Faraday tensor. \\ \hline \ 
Dual EM tensor & ${}^*\!F^{\mu\nu}$ & -- & Dual of electromagnetic tensor. \\ \hline \ 
Energy-momentum tensor & $T^{\mu\nu}$ & -- & Stress-energy tensor. \\ \hline \ 
Electric displacement & $D^i$ & $\frac{1}{\alpha}\left(E^i - \gamma^{ij} e_{jkl} \beta^k B^l\right)$ & Electric field measured by FIDO. \\ \hline \ 
Levi$-$Civita tensor & $e_{jkl}$ & $\epsilon_{ijk}\sqrt{\gamma}$ & Pseudo-tensor for cross products. \\ \hline \ 
Charge & $e$ & -- & Particle electric charge. \\ \hline \ 
Mass & $m$ & -- & Particle rest mass. \\ \hline \ 
Charge$-$to$-$mass ratio & $e/m$ & $10^8 \; C/kg$ & For protons. \\ \hline \ 
Gravitational radius & $r_g$ & $\frac{MG}{c^2}$ & Normalization for lengths. \\ \hline \ 
Larmor radius & $r_L$ & $\frac{\Gamma m v}{\alpha e B}$ & Radius of the gyro-orbit. \\ \hline \ 
Larmor frequency & $\omega_L$ & $\frac{\alpha e B}{2\pi \Gamma m}$ & Frequency of gyration. \\ \hline \ 
Energy at infinity & $E_\infty$ & $-u_0$ & Particle energy measured at infinity. \\ \hline \ 
Spin parameter &a & -- & Dimensionless BH spin ($0 \leq a < 1$). \\ \hline \ 
Gravitational time & $t_g$ & $r_g/c$ & Normalization for time. \\ \hline \ 
Escape time & $\tau_{\rm esc}$ & -- & Time for particle to leave outer boundary. \\ \hline \ 
Infall time & $\tau_{\rm in}$ & -- & Time for particle to fall into sink. \\ \hline \ 
Trapping time & $\tau_{\rm trap}$ & -- & Time particle spends in a cell. \\ \hline \ 
Diffusion coefficient & $\nu$ & $\frac{(\Delta R)^2 + (\Delta z)^2}{2 \eta \tau_{\rm trap}}$ & Characterizes diffusive transport. \\ \hline \ 
Power$-$law index & $\xi$ & -- & Index used to fit power$-$law distributions. \\ \hline \ 
Cylindrical radius & $R$ & -- & Radial coordinate in cylindrical system. \\ \hline \ 
Azimuthal angle & $\varphi$ & -- & Angular coordinate in cylindrical system. \\ \hline \ 
Vertical coordinate & $z$ & -- & Height coordinate in cylindrical system. \\ \hline \ 
Wavenumber & $\kappa_\ell$ & $L/\ell$ & Wavenumber at scale $\ell$ in coarse$-$graining. \\ \hline \ 
Scale & $\ell$ & -- & Coarse$-$graining scale. \\ \hline \ 
Black hole mass & $M$ & -- & Mass of central black hole. \\ \hline \ 
\end{tabular}
\caption{\label{table_symbols} Summary of all symbols and quantities appearing in the paper and their use.} 
\end{ruledtabular}
\end{table}

\end{document}